\documentclass{aa} 

\usepackage{graphicx}
\usepackage{txfonts,textcomp}

\usepackage{natbib}
\bibpunct{(}{)}{;}{a}{}{,}
\begin{document}

   \title{Facing the phase: Gravity-mode offset and buoyancy glitches in red--giant branch stars}

   \author{T. van Lier\inst{1}\fnmsep\inst{2}\fnmsep\thanks{Corresponding author; email: tobias.vanlier@h-its.org.}
        \and J. Müller\inst{1}\fnmsep\inst{2}
        \and S. Hekker\inst{1}\fnmsep\inst{2}
        }

   \institute{Heidelberg Institute for Theoretical Studies, Schloss-Wolfsbrunnenweg 35, D-69118 Heidelberg, Germany
            \and Zentrum für Astronomie Heidelberg (ZAH/LSW), Heidelberg University, Königstuhl 12, D-69117 Heidelberg, Germany}

   \date{Received 11 December 2025 / Accepted 15 May 2026}

    \abstract
   {With the increasing precision of asteroseismic observations, it becomes possible to reliably measure oscillation properties of an increasing number of stars. Interpreting these measurements requires a good theoretical understanding of their link to fundamental stellar properties.} 
   {In this study, we focus on the phase offset in gravity(g)-mode frequencies, which is imprinted in the asymptotic eigenfrequency pattern of mixed dipole modes observed in red--giant branch stars. We aim to unravel its physical origin and thus enable an informed interpretation of observations.}
   {Using stellar models, we empirically test the contribution of the g-mode offset $\varepsilon_\mathrm{g}$ (which is related to the wave reflection at cavity boundaries and commonly considered to be the dominant phase term) and glitches to the total observable phase.}
   {We find that, additionally to $\varepsilon_\mathrm{g}$, buoyancy glitches play an important role in the correct interpretation of the g--mode frequency phase. We further find that glitches in the evanescent zone also contribute to the phase, and we present a formalism to quantify this contribution. Finally, we propose a modification to the widely used formula for $\varepsilon_\mathrm{g}$.}
   {The g--mode frequency phase carries more information than previously considered. It has large analytic potential to study not only the reflection properties of the buoyancy cavity, but also the properties of glitches in the Brunt-Väisälä frequency.}


   \maketitle
   \nolinenumbers

\section{Introduction}\label{s:introduction}

The space-borne observation campaigns \textit{Kepler} and TESS have delivered long, high-precision timeseries of a large number of stars in the Milky Way. Among these are several ten thousand red-giant stars \citep[e.g.~][]{lit:Hekker2011,lit:Mosser2012,lit:Yu2018,lit:APOKASC1,lit:APOKASC2,lit:APOKASC3}. Thanks to their high luminosity, the sample of red giants bright enough to achieve a good signal-to-noise ratio in the light curves is also large, which makes statistical studies of their properties possible. Due to their large size and therefore cool surface temperatures, red-giant stars have convective envelopes, which stochastically drive solar-like oscillations: Energy from the turbulent convective motions is transferred to resonant global oscillation modes of the stars. This stochastic excitation leads to observable oscillations within a Gaussian power envelope centered around a frequency of maximum oscillation power $\nu_\mathrm{max}$, defined by the typical time scale of convection \citep{lit:Marthur2012}. It scales with a star's surface gravity $g$ and effective temperature $T_\mathrm{eff}$ as \citep[e.g.~][]{lit:Kjeldsen1995}
\begin{align}
    \nu_\mathrm{max}\propto\frac{g}{\sqrt{T_\mathrm{eff}}}\,.
    \label{eq:nu_max}
\end{align}

With advances in the automatization of asteroseismic analysis \citep[e.g.~][]{lit:Garcia2018,lit:Kallinger2019,lit:Corsaro2020,lit:Kuszlewicz2023}, an increasing number of targets is characterized in terms of oscillation properties. Asteroseismology uses the variability associated with perturbations to the stellar hydrostatic state -- for example the solar-like oscillations exhibited by red-giant stars -- to infer the properties of the equilibrium structure. A common approach to this analysis is to apply asymptotic theory, which uses a handful of parameters to characterize the spectrum of the stellar oscillations.

Each of these asymptotic parameters exists in two different contexts: Once as an empirical quantity used to reproduce the pattern of observed oscillation modes, and once as a theoretical quantity derived from the shape and behavior of a wave function. The latter is therefore determined by the fundamental stellar structure. For the asteroseismic observations to be useful, they need to be interpretable. Therefore, it is essential to match the empirical and theoretical counterparts. In this work, we investigate the phase term in the eigenfrequency pattern of mixed dipole modes observed in red--giant branch (RGB) stars. It is typically associated to the phase induced to the wave functions by partial reflection at turning points of the waves \citep[e.g.~][]{lit:Pincon2019}. \citet{lit:Cunha2015} showed that glitches in the propagation cavity -- sharp features in the stellar structure compared to the local wavelength -- can contribute an additional phase shift to the mode frequencies. The chemical discontinuity left behind by the first dredge-up creates a narrow spike in the Brunt-Väisälä frequency, which can act as such a glitch in RGB models. \citet{lit:Jiang2020} demonstrated that this spike affects the estimate of asymptotic oscillation parameters for the example of the coupling strength, for which they obtained unphysical values when the glitch was not considered. Further, \citet{lit:Jiang2022} were able to retrieve coupling strengths that matched the values expected from their models upon including the effects of the spike in the asymptotic expression for the eigenfrequencies. This motivates that considering the spike is also relevant to correctly interpret other asymptotic parameters, including the g--mode frequency phase.

Applying an observational approach to the eigenfrequencies of stellar models, we compare the empirical phase term with both theoretical contributions, the partial reflection at the turning points and the glitch. We also test different expansions and approximations that are commonly used for the theoretical calculations. In this way, we aim to gain understanding of the interpretation of observations, and also to enable more informed asymptotic estimates of eigenfrequencies for theoretical studies.

\section{Asymptotic oscillation theory}\label{s:asymptotics}

The fundamental equations of stellar oscillations can be derived by applying perturbations to the state variables in the stellar structure equations. Assuming spherical symmetry, the perturbations can be separated into a radial, angular and temporal part. The angular equations are solved by spherical harmonics of degree $\ell$ and azimuthal order $m$. Since we focus our analysis in this work on dipole modes, we implicitly set $\ell=1$ throughout. Since we further neglect rotation, magnetic fields, or any other non-isotropic effects, the order $m$ is irrelevant to all derivations in the following. \citet{lit:Takata2006b} showed that by asymptotic expansion (i.e.~assuming that the scale of the perturbations is small relative to the scale of variations in the equilibrium structure) the radial oscillation equations can be reduced to a single equation for the radial displacement relative to the center of mass $\xi_r$:
\begin{align}
    \frac{\mathrm{d}^2\xi_r}{\mathrm{d}r^2}=-k^2\xi_r\,. \label{eq:oscillation_equation}
\end{align}
The solutions for the other radial perturbations are proportional to the solution for $\xi_r$ up to a phase difference. The squared wave vector $k^2$ is linked to the angular frequency $\omega$, which appears in the solution to the temporal equations $\xi(t)\propto\mathrm{e}^{i\omega t}$, by the asymptotic dispersion relation:
\begin{align}
    k^2=\frac{\omega^2}{c_\mathrm{s}^2}\left(\left(\frac{\hat{S}}{\omega}\right)^2-1\right)\left(\left(\frac{\hat{N}}{\omega}\right)^2-1\right) \,.\label{eq:dispersion_relation}
\end{align}
Here, $c_\mathrm{s}$ is the local adiabatic sound speed. The characteristic Lamb ($\hat{S}$) and Brunt-Väisälä ($\hat{N}$) frequencies in this equation differ from their canonic forms by a factor $J$ introduced by the displacement of the center of mass \citep{lit:Takata2006b},
\begin{align}
    \hat{S}&=JS\,,\quad\hat{N}=\frac{N}{J}\,, \label{eq:SN_modification}\\
    J&=1-\frac{\rho}{\langle\rho\rangle_\mathrm{in}}\,,\label{eq:J_correction}
\end{align}
where $\rho$ is the local density of the stellar medium and $\langle\rho\rangle_\mathrm{in}$ is the average density of the stellar material below the radius where it is evaluated.

\begin{figure}[t]
    \resizebox{\hsize}{!}{\includegraphics[width=\textwidth]{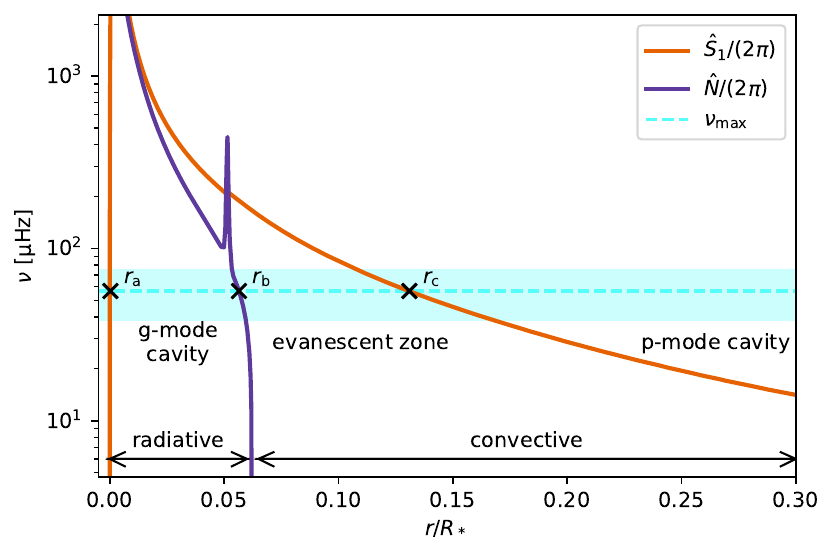}}
	\caption{Propagation diagram of the inner $30\,\%$ by radius of a red-giant model with $M=1.25\,\mathrm{M}_\odot,\,Z=0.020$. Characteristic frequencies $\hat{S}/(2\pi)$ (orange) and $\hat{N}/(2\pi)$ (purple) as a function of fractional radius. The dashed blue line and shaded area show the value of $\nu_\mathrm{max}$ and the range of dipole mode frequencies used in the analysis. In the g-mode cavity of a mode, its frequency lies below, and in the p-mode cavity, it lies above the two characteristic frequencies. Where $\hat{N}/(2\pi)<\nu<\hat{S}/(2\pi)$, the mode evanesces. Turning points at $\nu_\mathrm{max}$ are shown by crosses. The chemical discontinuity caused by the first dredge-up (i.e., the spike in $\hat{N}/(2\pi)$) is located just below the evanescent zone of the highest frequency considered in this model.}
	\label{fig:propagation_diagram}
\end{figure}

The solution to Eq.\,(\ref{eq:oscillation_equation}) is oscillatory with local wavelength $\frac{2\pi}{k}$ where $k^2>0$. This is the case for either $\omega^2<\hat{S}^2,\hat{N}^2$, in which case the oscillation is stabilized by buoyancy and corresponding modes are therefore referred to as buoyancy or gravity (g) modes, or for $\omega^2>\hat{S}^2,\hat{N}^2$, in which case these acoustic or pressure (p) modes propagate as sound waves. Where $k^2<0$, a perturbation decays exponentially with decay length $\kappa=ik\in\mathbb{R}$ according to Eq.\,(\ref{eq:oscillation_equation}). Oscillatory regions in a star are called cavities, regions in which a mode decays are called evanescent zones. In Fig.\,\ref{fig:propagation_diagram}, we show the propagation diagram (i.e., the local characteristic frequencies as a function of radius) for a typical RGB model. The internal g-mode cavity is bound by the modified Lamb frequency at the inner turning point $r_\mathrm{a}$ very close to the stellar center and by the modified Brunt-Väisälä frequency at the outer turning point $r_\mathrm{b}$. Above the evanescent zone, a p-mode cavity begins at $r_\mathrm{c}$ where the oscillation frequency crosses $\hat{S}$. The respective oscillation frequency of a mode is given by $\nu=\frac{\omega}{2\pi}$.

Imposing boundary conditions on the solutions of Eq.\,(\ref{eq:oscillation_equation}) states an eigenvalue problem with eigenvalues $\omega$. The resulting resonance frequencies of a star are at the heart of the asymptotic analysis. Their pattern in frequency space is what can be observed by a Fourier transform of the light curve.

\subsection{Eigenvalues of g modes}\label{ss:g-modes}

By expanding the dispersion relation Eq.\,(\ref{eq:dispersion_relation}) in the limit that $\omega\ll\hat{S},\hat{N}$ deep inside the buoyancy cavity, \citet{lit:Shibahashi1979} derived that pure g modes are evenly spaced in oscillation period $\Pi=\frac{2\pi}{\omega}$. The oscillation period of a pure g mode with radial order $n_\mathrm{g}$ ($n_\mathrm{g}<0$ by convention) is therefore given by
\begin{align}
    \Pi_{n_\mathrm{g}}^\mathrm{g}=(-n_\mathrm{g}+\Phi)\cdot\Delta\Pi\,, \label{eq:Pi_as_Mosser}
\end{align}
where $\Delta\Pi$ is the constant period spacing and $\Phi$ is some phase term. In the derivation by \citet{lit:Shibahashi1979}, $\Phi=0$. However, for example \citet{lit:Tassoul1980} suggested that the correct treatment of the wave function's behavior at the turning points would lead to a non-zero phase offset in Eq.\,(\ref{eq:Pi_as_Mosser}). \citet{lit:Mosser2012} empirically found that using $\Phi\not=0$ improved their asymptotic fits to observed mode-frequency patterns. In the literature, the empirical observed phase $\Phi$ and the theoretical g-mode offset due to partial reflection at the turning points are commonly identified with each other \citep[e.g.~][]{lit:Mosser2018,lit:Pincon2019}.

\subsection{Eigenvalues of mixed modes}\label{ss:mixed-modes}

As demonstrated by Fig.\,\ref{fig:propagation_diagram}, modes with frequencies around $\nu_\mathrm{max}$ can oscillate in both, the buoyancy cavity in the core and the acoustic cavity in the envelope of RGB stars. Oscillations in both cavities couple through the intermediate evanescent zone by the exchange of mode energy. As for the coupling of mechanical harmonic oscillators, this shifts the eigenfrequencies of the pure g and p modes to a joint eigenfrequency of the resulting mixed mode.

\citet{lit:Shibahashi1979} and \citet{lit:Takata2016b} showed that the eigenvalue condition for mixed modes can be written as
\begin{align}
    \cot(\Theta_\mathrm{g})\tan(\Theta_\mathrm{p})=q\label{eq:mixed_eigenvalue_condition}
\end{align}
in the framework of weak and strong interaction between the two cavities, respectively. In this expression, the coupling strength $q\in[0,1]$ represents the degree of interaction between the cavities, while $\Theta_\mathrm{g,p}$ quantify the deviation of the mixed mode frequency from the frequencies of the closest pure g and p mode, respectively \citep{lit:Jiang2018}.

Starting from Eq.\,(\ref{eq:mixed_eigenvalue_condition}) and using a formalism introduced by \citet{lit:CD2012}, \citet{lit:Hekker2018} derived an expression for the asymptotic oscillation periods of mixed modes\footnote{\citet{lit:Hekker2018} write $\arctan\left[q\cot(\Theta_\mathrm{p})\right]$ as a fre\-quen\-cy-de\-pen\-dent phase contributed by the coupling, which they call $\Phi(\nu)$.}:
\begin{align}
    \Pi_\mathrm{as}=\Delta\Pi\cdot\left(-n_\mathrm{pg}+\Phi-\frac{\arctan\left(q\cot(\Theta_\mathrm{p})\right)}{\pi}+n_\mathrm{p}\right)\,. \label{eq:Pi_as_mixed}
\end{align}
In this expression, $n_\mathrm{p}$ is the radial order of the pure p mode contributing to the mixed mode, and $n_\mathrm{pg}=n_\mathrm{p}+n_\mathrm{g}$ is the total radial order. Equation~(\ref{eq:Pi_as_mixed}) shows that the coupling of the pure g modes described by Eq.\,(\ref{eq:Pi_as_Mosser}) to a p mode introduces a phase shift to the mode periods. This phase is purely additive to $\Phi$. As $\Theta_\mathrm{p}$ (and hence the phase introduced by the mode coupling) is a function of frequency, it can be disentangled from the phase offset found for pure g modes (which is roughly constant across the observed frequency range, cf. e.g. \citet{lit:Pincon2019}), and therefore mixed modes can be used to study $\Phi$, even though it is a property of g modes. Equation~(\ref{eq:Pi_as_mixed}) is a function of the radial orders and frequencies of the mixed modes, along with five asymptotic parameters: the coupling strength~$q$, the g-mode parameters $\Delta\Pi$ and $\Phi$, and their p-mode equivalents -- the large frequency separation $\Delta\nu$ and p--mode phase offset $\varepsilon_\mathrm{p}$ -- which are hidden in $\Theta_\mathrm{p}$,
\begin{align}
    \Theta_\mathrm{p}=\pi\left(\frac{\nu}{\Delta\nu}-(n_\mathrm{p}+\varepsilon_\mathrm{p})\right)\,.
\label{eq:Theta_p}
\end{align}

\section{Gravity-mode offset $\varepsilon_\mathrm{g}$}\label{s:epsg}

At the turning points of cavities, propagating waves are partially reflected, which induces a phase shift to the stand\-ing-wave frequencies observed as eigenfrequencies of the star. For the g-mode component of oscillations in red giants, \citet{lit:Takata2016a} derived the phase contributed by reflection at the inner turning point to be $\frac{1}{2}$ because the modified Lamb frequency $\hat{S}$ is very steep around $r_\mathrm{a}$ (see Fig.\,\ref{fig:propagation_diagram}). The phase shift at the outer turning point, collected into a term $\varepsilon_\mathrm{g}$, is given by
\begin{align}
    \varepsilon_\mathrm{g}=-\frac{1}{\pi}\frac{\mathrm{d}}{\mathrm{d}\nu}\int_{\tilde{r}}^{r_\mathrm{b}}\nu\cdot k(r,\nu)\,\mathrm{d}r+\frac{\Psi}{\pi} \label{eq:epsg_Takata}
\end{align}
in the framework of strong interaction between the oscillation cavities introduced by \citet{lit:Takata2016a}. The radius coordinate $\tilde{r}$ is chosen deep inside the buoyancy cavity. In this expression for $\varepsilon_\mathrm{g}$ \citep[inspired by the analysis by][]{lit:Pincon2019}, the first term corresponds to the phase contribution by reflection at $r_\mathrm{b}$ alone, while $\Psi$ represents the interaction with the sound cavity across a narrow evanescent zone. Both terms are discussed in further detail in the following subsections.

\subsection{Numerical evaluation of the reflection at $r_\mathrm{b}$}\label{ss:epsg_numeric}

Before numerically evaluating the first term on the right-hand side of Eq.\,(\ref{eq:epsg_Takata}), we simplified it analytically in order to reduce the numerical noise. Since $r_\mathrm{b}$ depends on the oscillation frequency, we use that
\begin{align}
    \frac{\partial}{\partial y}\int_{a}^{g(y)}f(x,y)\,\mathrm{d}x=\int_{a}^{g(y)}\frac{\partial f(x,y)}{\partial y}\,\mathrm{d}x+\frac{\partial g(y)}{\partial y}\cdot f(g(y),y) \label{eq:der_int}\,,
\end{align}
which we can apply to the expression at hand by identifying $x\equiv r$, $y\equiv\nu$, $g(y)\equiv r_\mathrm{b}$, and $f(x,y)\equiv\nu k(r,\nu)$. Because the oscillation frequency is constant across the radius of the star, $\frac{\partial\nu}{\partial r}=0$, we can promote Eq.\,(\ref{eq:der_int}) from partial to total derivatives in this case. We therefore obtain
\begin{align}
    \frac{\mathrm{d}}{\mathrm{d}\nu}\int_{\tilde{r}}^{r_\mathrm{b}}\nu k(r,\nu)\,\mathrm{d}r = \int_{\tilde{r}}^{r_\mathrm{b}}\frac{\partial[\nu\cdot k(r,\nu)]}{\partial\nu}\,\mathrm{d}r + \frac{\partial r_\mathrm{b}}{\partial\nu}\cdot\nu\cdot k(r_\mathrm{b},\nu)\,. \label{eq:der_int_applied}
\end{align}
The second term on the right-hand side of this equation is equal to zero since $k(r_\mathrm{b},\nu)=0$ by the definition of $r_\mathrm{b}$ and the other factors are finite. This leaves only one term. The partial derivative can be evaluated using the asymptotic dispersion relation Eq.\,(\ref{eq:dispersion_relation}). We obtain
\begin{align}
    \frac{\partial[\nu\cdot k(r,\nu)]}{\partial\nu} = k\cdot\left(2-\frac{1}{1-\left(\frac{2\pi\nu}{\hat{S}}\right)^2}-\frac{1}{1-\left(\frac{2\pi\nu}{\hat{N}}\right)^2}\right)\,. \label{eq:epsg_integrand}
\end{align}

\subsection{Evanescent zone contribution $\Psi$}\label{ss:Psi}

\citet{lit:Takata2016a} finds the following expression for the phase contributed to the g-mode frequencies by the interaction with the sound cavity across the evanescent zone:
\begin{align}
    \Psi&=\Psi_0(X)-\frac{1}{2}\arg[\mathsf{D}(0)]-\frac{\pi}{4}\,, \label{eq:Psi_Takata}\\
    \Psi_0&=-\frac{1}{2}\arg[\Gamma(1+iX)]-\frac{X}{2}(1-\ln(X))+\frac{\pi}{8}\,. \label{eq:Psi0_Takata}
\end{align}
Here, $\arg$ and $\Gamma$ are the complex argument and gamma functions, respectively, and $\mathsf{D}(0)$ and $X$ are functions of the stellar structure in the evanescent zone, notably containing gradients of the characteristic frequencies evaluated at $\sqrt{r_\mathrm{b}r_\mathrm{c}}$. For the full form of the expressions, we refer the reader to \citet{lit:Takata2016a}. We note that $\arg[\Gamma(1+iX)]$ is chosen to be the standard argument with values in $[-\pi,\pi]$ around $X=0$, and then offset by multiples of $2\pi$ to maintain a smooth function for large $X$.

Crucially, for wide evanescent zones, $X$ goes to infinity and therefore $\Psi_0$ goes to zero, while $\mathsf{D}(0)$ becomes real and hence $\arg[\mathsf{D}(0)]$ also goes to zero. Therefore
\begin{align}
    \Psi \xrightarrow{X\rightarrow\infty}-\frac{\pi}{4} \label{eq:Psi_lim}
\end{align}
in the limit of a wide evanescent zone. This expansion is commonly used in the literature when calculating $\varepsilon_\mathrm{g}$, and on comparison to observations, it appears to give decent results \citep[e.g.~][]{lit:Pincon2019}. From a theoretical point of view, however, it is somewhat inconsistent to use. On the one hand, in early red giants \citep[when the evanescent zone is narrow,][]{lit:vanRossem2024} it is not applicable because the requirement of a wide evanescent zone is clearly not met. We will investigate how strong the deviation of the full expression from this limit is, and whether it is in agreement with both observations and models. On the other hand, as the evanescent zone widens, the underlying assumptions leading to the expression in Eq.\,(\ref{eq:Psi_Takata}) are no longer met, since it was itself derived in the limit of a narrow evanescent zone. As shown by, for example, \citet{lit:vanLier2025}, the results derived by \citet{lit:Takata2016a} in this limit are only valid on the early RGB. It is plausible, however, that as the interaction with the sound cavity becomes weaker the evanescent-zone properties contribute less to the phase shift imprinted on the g-mode frequencies and the terms in $X$ vanish. Therefore, it seems reasonable to assume that in this case the limit still holds. We will also test this assumption.

When evaluating the full expression for $\Psi$, Eq.\,(\ref{eq:Psi_Takata}), for our stellar models, we find a strong discrepancy between our results and the values obtained by \citet{lit:Takata2016a} for an analytical model of the stellar structure. Most notably, we obtain significantly higher values for $\arg[\mathsf{D}(0)]$, leading to substantially lower $\varepsilon_\mathrm{g}$. These do not agree well with observed g-mode offsets, while agreement is increased when omitting that term altogether. The term $\arg[\mathsf{D}(0)]$ appears as the antiderivative at the lower limit of an integral in Eq.\,(A16) of \citet{lit:Takata2016a}.\footnote{In the form written in this equation, $\tilde{\mathsf{D}}$ is the function $\mathsf{D}$ scaled by a large quantity such that the absolute value of $\tilde{\mathsf{D}}$ is of order unity. Then \begin{align*}\ln\left(\tilde{\mathsf{D}}(0)\right)=\ln\left(|\tilde{\mathsf{D}}(0)|\mathrm{e}^{i\arg[\tilde{\mathsf{D}}(0)]}\right)=\ln\left(|\tilde{\mathsf{D}}(0)|\right)+i\arg[\tilde{\mathsf{D}}(0)]\approx i\arg[\mathsf{D}(0)]\,,\end{align*} because $\ln(1)=0$ and the argument of a complex number is independent of scaling.} The evaluation of the corresponding integral is presented in Appendix~A.1.3 of the same paper, and it crucially depends on Eq.\,(A41) thereof, where it is assumed that the inverse of the coordinate value at the upper limit of the integral $|\mathfrak{x}|^{-1}$ is of the same order of magnitude or smaller than the inverse of the coordinate value at the edges of the evanescent zone $|x_0|^{-1}$. The coordinates $x$ are defined in such a way that $0\leq|x|\leq|x_0|$ in the evanescent zone, where $x=0$ in the center of the evanescent zone at $r=\sqrt{r_\mathrm{b}r_\mathrm{c}}$.
Since the integral is evaluated with the upper limit $x=\mathfrak{x}$ somewhere within the evanescent zone as part of the wave function at $\mathfrak{x}$ (which is why only the lower limit of the integral is a constant that is absorbed into the phase $\Psi$), $|\mathfrak{x}|^{-1}\geq|x_0|^{-1}$. For some values of $\mathfrak{x}$, the assumption made in Eq.\,(A41) of \citet{lit:Takata2016a} will therefore not hold. The full evaluation of the integral for arbitrary $\mathfrak{x}$ is not straightforward. Taking the opposite limit, however, and choosing $|x|\approx0$ everywhere in the integration domain makes part of the integrand vanish, which removes the term $\arg[\mathsf{D}(0)]$ from the phase entirely, so we are only left with
\begin{align}
    \hat{\Psi} = \Psi_0-\frac{\pi}{4}\,, \label{eq:Psi_new}
\end{align}
where $\Psi_0$ is given by Eq.\,(\ref{eq:Psi0_Takata}).\footnote{This also implies that the antiderivative at the upper limit of the integral $\propto\ln(\mathsf{D}(s))$ is missing in the wave function in this limit. It goes beyond the scope of this work to follow through on the full impact of this assumption.} This limit is equivalent to assuming that the contribution from the middle of the evanescent zone dominates the integral.

\subsection{An analytical model for the buoyancy cavity}\label{ss:Pincon_model}

In an effort to evaluate $\varepsilon_\mathrm{g}$ fully analytically, \citet{lit:Pincon2019} developed a simplified model of the Brunt-Väisälä frequency and assumed $\hat{S}\gg2\pi\nu$ for $\tilde{r}\leq r\leq r_\mathrm{b}$. In this toy model, the profile of $\hat{N}$ is separated into two regimes, following a power law inside the buoyancy cavity and abruptly dropping to zero at the bottom of the convective envelope,
\begin{align}
    \hat{N}(r)=\left\{
    \begin{array}{ll}
         \hat{N}_\mathrm{BCZ}\left(\frac{r_\mathrm{BCZ}}{r}\right)^\beta\,, & \tilde{r}\leq r\leq r_\mathrm{BCZ}  \\
         0\,, & r_\mathrm{BCZ}<r
    \end{array}
    \right.\,. \label{eq:N_Pincon}
\end{align}
The subscript \lq BCZ\rq\ refers to quantities at the bottom of the convection zone. The authors further assumed the modifying factor~$J$ (Eq.\,\ref{eq:J_correction}) to be constant due to the high density contrast between stellar core and envelope. In this case, it relates to the power-law index $\beta$ as $J=\frac{2\beta}{3}$.

Under these assumptions, the g-mode offset can be calculated analytically for two different cases. For $2\pi\nu>\hat{N}_\mathrm{BCZ}$ (i.e., the buoyancy cavity does not extend all the way to the convective envelope and $r_\mathrm{b}$ is a function of frequency), the variations of $k\nu$ and $r_\mathrm{b}$ with oscillation frequency cancel and $\varepsilon_\mathrm{g}$ becomes independent of $\nu$. \citet{lit:Pincon2019} refer to this as \lq case a\rq:
\begin{align}
    \varepsilon_\mathrm{g}^\mathrm{a}=\frac{\sqrt{2}}{3}-\frac{1}{4}\,,\qquad2\pi\nu>\hat{N}_\mathrm{BCZ}\,, \label{eq:epsg_Pincon_a}
\end{align}
where we use the limit $\Psi\approx-\frac{\pi}{4}$ for wide evanescent zones for consistency with the assumption that $\hat{S}\gg2\pi\nu$. In \lq case b\rq\ ($2\pi\nu<\hat{N}_\mathrm{BCZ}$) the turning point stays fixed at $r_\mathrm{BCZ}$ when varying the oscillation frequency, and the induced phase becomes frequency-dependent \citep{lit:Pincon2019}:
\begin{align}
    \varepsilon_\mathrm{g}^\mathrm{b}=\frac{2\sqrt{2}}{3\pi}\arcsin\left(\frac{2\pi\nu}{\hat{N}_\mathrm{BCZ}}\right)-\frac{1}{4}\,,\qquad2\pi\nu<\hat{N}_\mathrm{BCZ}\,. \label{eq:epsg_Pincon_b}
\end{align}
During evolution along the RGB, the frequency range in which oscillations are excited decreases as the envelope expands. At the same time, as the core contracts, $\hat{N}_\mathrm{BCZ}$ increases. This means that a red giant will evolve from case a to case b as it ages.

\section{Glitches in the Brunt-Väisälä frequency}\label{s:glitches}

Glitches are sharp variations in the stellar structure compared to the local wavelength of the oscillations. In RGB stars, the spike in the Brunt-Väisälä frequency (cf.~Fig.\,\ref{fig:propagation_diagram}) representing the chemical discontinuity at the deepest extent of the convective envelope during the first dredge-up is the most prominent sharp feature that can act as a glitch. During evolution, it first appears in the evanescent zone of excited mixed modes and moves into the buoyancy cavity as $\nu_\mathrm{max}$ decreases \citep[cf.~also the discussion by][]{lit:vanLier2025}.

\subsection{A Dirac-delta glitch in the evanescent zone}\label{ss:glitch_EZ}

Typically, glitches are understood as narrow features within the oscillation cavities of a star. \citet{lit:Jiang2022} showed that the spike in the Brunt-Väisälä frequency affects the inferred coupling strength also when it lies in the evanescent zone of the dipole modes. Here, we derive the g--mode frequency phase contributed by a glitch in the evanescent zone. We start our derivation from the formalism introduced by \citet{lit:Shibahashi1979}. This means we implicitly assume weak coupling between the cavities (which we do as \citet{lit:vanLier2025} have shown that the strong-coupling formalism introduced by \citet{lit:Takata2016a} is only valid for less evolved models for which the spike does not appear in the evanescent zone). In the case of weak coupling, the interaction is purely additive in the wave functions. This means that the coupling only adds a (frequency-dependent) phase to the eigenfrequencies. We further assume that the glitch is far enough away from the p-mode cavity not to affect the p-mode eigenfrequencies, so the g-mode phase contributed by the coupling ($\arctan\left[q\cot(\Theta_\mathrm{p})\right]/\pi$, cf. Eq.\,\ref{eq:Pi_as_mixed}) is independent of the presence of the glitch.
Also, we neglect the phase contributed by the behavior of the wave function around the turning points $r_\mathrm{a}$ and $r_\mathrm{b}$. Since we are only interested in the relative phase contributed by the glitch, both $\varepsilon_\mathrm{g}$ and the effect of the (weak) coupling cancel out when taking the difference between the eigenvalue condition with and without a spike. This allows us to apply this simplified framework, in which we can treat the g-mode component of the mixed modes as isolated g~modes. We note, however, that this framework does not allow for the inclusion of any potential interactions between the phases contributed by the coupling and the glitch in the evanescent zone.

\begin{figure}[t]
    \resizebox{\hsize}{!}{\includegraphics[width=\textwidth]{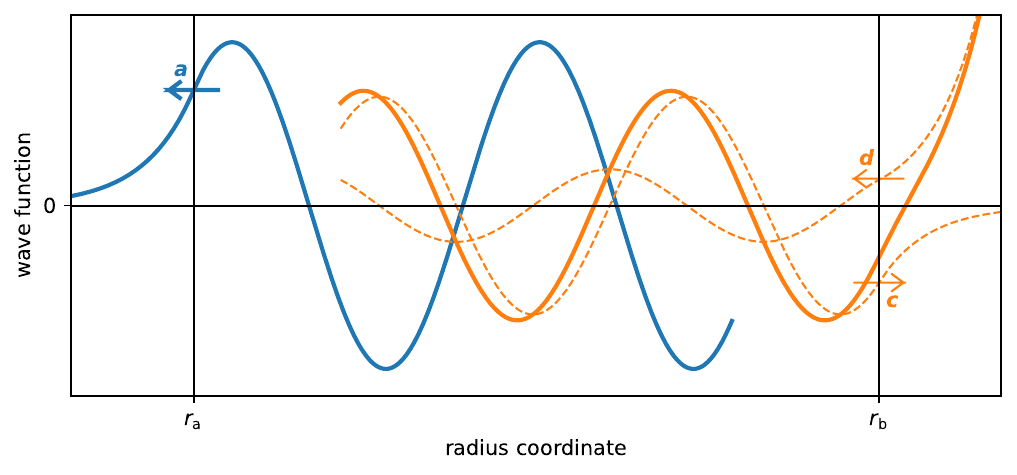}}
	\caption{Sketch of the setup of wave functions used in Sect.\,\ref{ss:glitch_EZ}, with wave functions labeled by their amplitude parameters. The wave function shown as a solid orange line is the sum of the two dashed orange lines. When deriving the eigenfrequencies, the solid blue and orange lines are matched deep inside the cavity (i.e., between $r_\mathrm{a}$ and $r_\mathrm{b}$).}
	\label{fig:Shibahashi_sketch}
\end{figure}

The eigenvalue condition for g~modes is obtained by matching two asymptotic solutions to the oscillation equations that are expanded around the different turning points. The first solution describes an outgoing wave to the center of the star with an amplitude $a$, originating from an expansion around the inner turning point $r_\mathrm{a}$. The second solution is a superposition of an outgoing wave to and an incoming wave from the stellar envelope with amplitudes $c$ and $d$, respectively, originating from an expansion around the outer turning point $r_\mathrm{b}$. This setup is sketched in Fig.\,\ref{fig:Shibahashi_sketch}. The amplitude of an incoming wave from the stellar center has to be $b\equiv0$ to ensure regularity of the solution at $r=0$.

Imposing that the solutions around $r_\mathrm{a}$ and $r_\mathrm{b}$ match deep inside the buoyancy cavity for two dependent variables (representing the radial displacement and pressure perturbation, the latter of which is essentially a radial derivative of the former), we arrive at a system of equations for the amplitudes:
\begin{align}
    c&=-a\cdot\sin\left(\int_{r_\mathrm{a}}^{r_\mathrm{b}}k\,\mathrm{d}r-\frac{\pi}{2}\right)\,,\label{eq:Shibahashi_c}\\
    d&=+a\cdot\cos\left(\int_{r_\mathrm{a}}^{r_\mathrm{b}}k\,\mathrm{d}r-\frac{\pi}{2}\right)\,.\label{eq:Shibahashi_d}
\end{align}
We note that the pure g~modes considered by \citet{lit:Shibahashi1979} are different from those studied here since the nature of the turning points is exchanged. This introduces a phase shift of $\pi$ to the amplitude relation for $d$, hence Eq.\,(\ref{eq:Shibahashi_d}) differs from Eq.\,(21) in \citet{lit:Shibahashi1979} by a `$-$' sign. Starting from these amplitude relations, \citet{lit:Shibahashi1979} argued that the incoming wave from the envelope should have a vanishing amplitude ($d=0$) since otherwise the solution would grow exponentially towards the surface of the star. Using Eq.\,(\ref{eq:Shibahashi_d}), this leads to the eigenvalue condition
\begin{align}
    \cos\left(\int_{r_a}^{r_b}k\,\mathrm{d}r-\frac{\pi}{2}\right)=0 \quad\Leftrightarrow\quad \int_{r_a}^{r_b}k\,\mathrm{d}r=n\pi \label{eq:Shibahashi_eigenvalue}
\end{align}
for some integer $n$. As this condition is periodic in $\pi$, we retrieve the same equation as in the original paper despite the phase shift due to the turning-point nature.

We now assume that the glitch in the evanescent zone can lead to a reflection of mode energy. This allows $d$ to become non-zero without the need for a growing mode beyond the glitch. Mathematically, we parametrize this using the following shape of a generic `wave' function in an evanescent environment:
\begin{align}
    \psi(r)=\left\{
    \begin{array}{ll}
         \exp\left(-\int_{r_\mathrm{b}}^{r}\kappa_0\mathrm{d}r\right)+R\exp\left(+\int_{r_\mathrm{b}}^{r}\kappa_0\mathrm{d}r\right)\,,&r_\mathrm{b}\leq r<r^* \\
          T\exp\left(-\int_{r_\mathrm{b}}^{r}\kappa_0\mathrm{d}r\right)\,,&r>r^*
    \end{array}
    \right.\,, \label{eq:psi_EZ_glitch}
\end{align}
where $r^*$ indicates the radial position of the glitch, which we model as a Dirac-delta function superimposed on a smooth squared decay length $\kappa_0^2(r)$. This representation of $\psi(r)$ relies on the assumption that the glitch is located sufficiently far away from the turning point that the wave function takes its asymptotic form at $r^*$. $R$ and $T$ are amplitude coefficients corresponding to reflection and transmission at the glitch, respectively. This means that we can identify $d$ in the framework of \citet{lit:Shibahashi1979} as
\begin{align}
    d=Rc\,.\label{eq:R_to_d}
\end{align}

We now require $\psi$ to be continuous at $r^*$,
\begin{align}
    \lim_{r\nearrow r^*}\psi=\lim_{r\searrow r^*}\psi=:\psi(r^*)\,,
    \label{eq:psi_continuity}
\end{align}
which relates the reflection and transmission coefficients:
\begin{align}
    R=(T-1)\cdot\mathrm{e}^{-2\mathfrak{d}^*}\,,
    \label{eq:RT_relation}
\end{align}
where $\mathfrak{d}^*$ denotes the decay of the mode amplitude up until the glitch in number of e-foldings,
\begin{align}
    \mathfrak{d}^*=\int_{r_\mathrm{b}}^{r^*}\kappa_0\,\mathrm{d}r\,.
    \label{eq:decstar}
\end{align}
The radial derivative of the wave function $\frac{\mathrm{d}\psi}{\mathrm{d}r}$ cannot be expected to be continuous in the presence of a glitch. Instead, we use a narrow integral of the oscillator equation~(\ref{eq:oscillation_equation}) around $r^*$ to quantify the discontinuity:
\begin{align}
    \lim_{\epsilon\rightarrow0}\int_{r^*-\epsilon}^{r^*+\epsilon}\frac{\mathrm{d}^2\psi}{\mathrm{d}r^2}=-\lim_{\epsilon\rightarrow0}\int_{r^*-\epsilon}^{r^*+\epsilon}k^2\psi\,\mathrm{d}r\,.
    \label{eq:osci_equation_integral}
\end{align}
Assuming a Dirac-delta function for the glitch, this limit is equivalent to
\begin{align}
    \lim_{r\searrow r^*}\frac{\mathrm{d}\psi}{\mathrm{d}r}-\lim_{r\nearrow r^*}\frac{\mathrm{d}\psi}{\mathrm{d}r}=-A\cdot\psi(r^*)\,,
    \label{eq:dpsi_dr_constraint}
\end{align}
where $A$ is the amplitude of the glitch in the sense of the area under the spike,
\begin{align}
    A=\int\limits_\mathrm{glitch}(k^2-k_0^2)\,\mathrm{d}r\,.
    \label{eq:EZ_amplitude}
\end{align}
Using Eq.\,(\ref{eq:psi_EZ_glitch}) and neglecting derivatives of the decay length, this leads to
\begin{align}
    \kappa_0(r^*)\cdot\left(-T\mathrm{e}^{-\mathfrak{d}^*}+\mathrm{e}^{-\mathfrak{d}^*}-R\mathrm{e}^{+\mathfrak{d}^*}\right)=-A\cdot T\mathrm{e}^{-\mathfrak{d}^*}\,.
    \label{eq:divpsi_dr_result}
\end{align}
Combining Eqs.\,(\ref{eq:RT_relation}) and~(\ref{eq:divpsi_dr_result}), we obtain the reflection coefficient
\begin{align}
    R=\frac{1}{\mathfrak{A}^{-1}-1}\cdot\mathrm{e}^{-2\mathfrak{d}^*}\,,
    \label{eq:reflection_coefficient}
\end{align}
with the scaled amplitude parameter
\begin{align}
    \mathfrak{A}=\frac{A}{2\kappa_0(r^*)}\,.
    \label{eq:glitchA}
\end{align}

Inserting Eq.\,(\ref{eq:reflection_coefficient}) into~(\ref{eq:R_to_d}) and further into (\ref{eq:Shibahashi_d}), together with Eq.\,(\ref{eq:Shibahashi_c}) this leads to a new eigenvalue condition:
\begin{align}
    &\cos\left(\int_{r_\mathrm{a}}^{r_\mathrm{b}}k\,\mathrm{d}r-\frac{\pi}{2}\right)+R\sin\left(\int_{r_\mathrm{a}}^{r_\mathrm{b}}k\,\mathrm{d}r-\frac{\pi}{2}\right)=0 \\
    &\Leftrightarrow \int_{r_\mathrm{a}}^{r_\mathrm{b}}k\,\mathrm{d}r=\left(n+\left[\frac{1}{\pi}\arctan\left((1-\mathfrak{A}^{-1})\cdot\mathrm{e}^{+2\mathfrak{d}^*}\right)+\frac{1}{2}\right]\right)\pi\,. \label{eq:EZ_glitch_eigenvalue}
\end{align}
We call the phase term in square brackets, which is added to the original condition Eq.\,(\ref{eq:Shibahashi_eigenvalue}) due to the presence of the glitch,
\begin{align}
    \Phi_\mathrm{EZ}\coloneqq\frac{1}{\pi}\arctan\left((1-\mathfrak{A}^{-1})\cdot\mathrm{e}^{+2\mathfrak{d}^*}\right)+\frac{1}{2}\,. \label{eq:Phi_EZ}
\end{align}
For $\mathfrak{A}<1$ (which was the case in all models we analyzed), $\Phi_\mathrm{EZ}$ is limited from above by $\frac{1}{2}$ and goes to zero for large distances from the turning point to the glitch $r^*-r_\mathrm{b}$ (equivalent to large $\mathfrak{d}^*$) or small glitch amplitudes $\mathfrak{A}$. We note that for very large glitches ($\mathfrak{A}>1$), $\Phi_\mathrm{EZ}$ is limited by $\frac{1}{2}$ from below and approaches 1 in the limit of the glitch being very far away from $r_\mathrm{b}$. Since the g--mode frequency phase is only defined modulo~1, the limit of a zero effective phase contribution for large $\mathfrak{d}^*$ is maintained in this case, too.

\subsection{A Dirac-delta glitch in the g-mode cavity}\label{ss:glitch_cavity}

The phase contribution to the eigenfrequencies of mixed modes by a spike-like glitch in the buoyancy cavity was first derived by \citet{lit:Cunha2015}. The glitch discussed here is inherently a feature of the Brunt-Väisälä frequency. When deriving the phase it contributes to the eigenfunction of an oscillation mode, however, the sharp variation that the spike causes in the corresponding wave vector is relevant (similarly as in Eq.\,29). In the limit $\omega^2\ll\hat{N}^2,\hat{S}^2$, which is valid deep inside the buoyancy cavity, $k^2\approx \frac{2N^2}{\omega^2r^2}$ for dipole modes. Therefore, the amplitude of the resulting spike in $k^2$ is proportional to that in $N^2$ and so it can be parametrized by a frequency-independent amplitude multiplied by $\omega^{-2}$ in this limit. This approach is taken by \citet{lit:Cunha2015}.

They modeled the glitch as a Dirac-delta perturbation with amplitude $C$ to a smooth squared Brunt-Väisälä frequency $N_0^2$,
\begin{align}
    N^2(r)=N_0^2(r)\cdot\left(1+C\delta(r-r^*)\right)\,.
    \label{eq:delta_N}
\end{align}
Using the expansion for the wave vector discussed above, they derived resonance frequencies in the framework developed by \citet{lit:Gough1993}. They continuously matched asymptotic wave functions expanded around $r_\mathrm{a}$ and $r_\mathrm{b}$ at $r^*$, and quantified the jump in the radial derivative of the wave function at $r^*$ analogously to Eq.\,(\ref{eq:dpsi_dr_constraint}), where $A$ needs to be replaced by $\frac{2N_0^2(r^*)}{\omega^2{r^*}^2}C$. Comparing the resulting eigenfrequencies to those derived without a glitch, \citet{lit:Cunha2015} found that a spike in $N^2$ introduces a phase shift of $\Phi_{\delta}$ given by\footnote{We note that $\Phi_{\delta}$ here differs from $\Phi$ as derived by \citet{lit:Cunha2015} by a factor $\pi$, because \citet{lit:Cunha2015} interpret it as a phase to the wave function rather than the eigenfrequencies.}
\begin{align}
    \Phi_{\delta}&=\frac{1}{\pi}\arcsin\left[\frac{\mathfrak{C}}{\omega B}\sin^2\left(\phi+\frac{\pi}{4}\right)\right]\,, \label{eq:Phi_Cunha}\\
    B^2&=\left[1-\frac{\mathfrak{C}}{2\omega}\cos\left(2\phi\right)\right]^2 + \left[\frac{\mathfrak{C}}{\omega}\sin^2\left(\phi+\frac{\pi}{4}\right)\right]^2\,, \label{eq:B_Cunha}\\
    \mathfrak{C}&=\frac{\sqrt{2}N_0(r^*)}{r^*}C\,. \label{eq:glitchC}
\end{align}
Here, $\phi$ characterizes the position of the glitch as the phase of the eigenfunction at $r^*$. This phase is comprised of the phase the eigenfunction picks up when propagating from $r_\mathrm{b}$ to $r^*$, approximated by an integral over the asymptotic wave vector, and the phase shift induced by the coupling to the p-mode cavity \citep{lit:Cunha2015,lit:Cunha2024}:
\begin{align}
    \phi&=\int_{r^*}^{r_\mathrm{b}}k_0\mathrm{d}r+\arctan\left(q\cot(\Theta_\mathrm{p})\right)\,.\label{eq:phase_position}
\end{align}
Applying the discussed limit for the smooth wave vector $k_{0}$ and correcting for the behavior close to the turning point by introducing a phase offset $\delta$, the first term can be substituted by \citep{lit:Cunha2019}:
\begin{align}
    \int_{r^*}^{r_\mathrm{b}}k_0\mathrm{d}r\approx\int_{r^*}^{r_\mathrm{b}}\frac{\sqrt{2}N_0}{\omega r}\,\mathrm{d}r+\delta =:\frac{\omega_\mathrm{g}^*}{\omega}+\delta\,.\label{eq:omgstar}
\end{align}
The parameter $\omega_\mathrm{g}^*$ is the buoyancy depth of the glitch.

Following the arguments in Sect.\,\ref{s:epsg}, $\varepsilon_\mathrm{g}$ represents the phase in the resonance frequencies due to corrections close to $r_\mathrm{b}$, equivalent to a phase $\pi\varepsilon_\mathrm{g}$ imparted on the eigenfunction in this region. While these corrections do not only account for deviations from the limit $k^2\propto\omega^{-2}$, but also for corrections of the non-asymptotic behavior of the wave function around $r_\mathrm{b}$, and are hence conceptually different from $\delta$ as it is introduced in the approximation above, we argue that they should be taken into account in the calculation of the phase of the eigenfunction at $r^*$. We therefore use the mathematically similar expression:
\begin{align}
    \phi=\frac{\omega_\mathrm{g}^*}{\omega}+\pi\varepsilon_\mathrm{g,as}+\arctan\left(q\cot(\Theta_\mathrm{p})\right)\,,\label{eq:delta}
\end{align}
to estimate $\phi$. However, we note that this constitutes an additional assumption about the phase $\delta$, which is left free in the formulation by \citet{lit:Cunha2024}. The subscript `as' indicates that Eq.\,(\ref{eq:delta}) refers to the g-mode offset found from the asymptotic frequency pattern.

The glitch derived in this framework contributes a negative phase to the eigenfrequencies, so $\Phi_{\delta} >0$ is subtracted from the integer $n$ in the eigenvalue condition of the g~modes (Eq.\,\ref{eq:Shibahashi_eigenvalue}). We note that, analogously to Sect.\,\ref{ss:glitch_EZ}, this derivation assumes that the wave function takes its asymptotic form either side of $r^*$ or, equivalently, that the glitch is located sufficiently deep inside the buoyancy cavity.

\subsection{A Gaussian glitch in the g-mode cavity}\label{ss:glitch_GaussCunha}

Another assumption inherent to both glitch parametrizations presented thus far is that the spike in the Brunt-Väisälä frequency can be modeled by a Dirac-delta function. This is only accurate for a step-like discontinuity in the chemical abundance profile of the star producing a Dirac-delta spike in the Brunt-Väisälä frequency. During evolution, diffusive mixing processes will successively smooth the sharp transition, making the width of the spike finite. As the star ages, the glitch moves further into the cavity and to significantly higher wavenumbers $k_0(r^*)$, further increasing its width relative to the local wavelength $\frac{2\pi}{k_0}$. When the glitch becomes too wide, the approximation by a Dirac-delta function is no longer suitable. \citet{lit:Cunha2019,lit:Cunha2024} presented a term for the phase contributed by a glitch in $N$ modeled by a Gaussian function,
\begin{align}
    N(\omega_\mathrm{g}^{r})=N_0\cdot\left(1+\frac{C_\mathrm{G}}{\sqrt{2\pi}\Delta_\mathrm{g}}\exp\left(-\frac{(\omega_\mathrm{g}^{r}-\omega_\mathrm{g}^*)^2}{2\Delta_\mathrm{g}^2}\right)\right)\,,
\end{align}
instead. Here, $\omega_\mathrm{g}^{r}$ is the buoyancy depth corresponding to the radius coordinate (cf.~Eq.\,(\ref{eq:omgstar}), $\omega_\mathrm{g}^*=\omega_\mathrm{g}^{r}(r^*)$), $\Delta_\mathrm{g}$ is the width of the glitch parametrized by the standard deviation of the Gaussian in units of buoyancy depth, and $C_\mathrm{G}$ is the amplitude of the Gaussian.

\citet{lit:Cunha2019} used a similar approach to \citet{lit:Cunha2015} to derive the glitch-induced phase shift, now for a Gaussian instead of a Dirac-delta glitch. The challenge, however, is that for a finite width of the Gaussian, they needed to assume a shape of the wave function inside the glitch. This is not straightforward, since the asymptotic expansion and the whole analysis starting from Eq.\,(\ref{eq:oscillation_equation}) are not valid in the regime of fast variations in stellar structure. \citet{lit:Cunha2019} tried out different formulations and introduced some factors ad-hoc to both match numerically obtained phases and converge to different limits correctly, which in the end left them with an expression that is not rigorously derived but motivated by the mathematical formalism and supported by numeric results. \citet{lit:Cunha2024} summarized the obtained phase contribution:
\begin{align}
    \Phi_\mathrm{G}&=\frac{1}{\pi}\mathrm{arccot}\left[\frac{1}{C_\mathrm{G}f_\omega^{\Delta_\mathrm{g}}\sin^2(\beta_2)}-\cot(\beta_2)\right]\,,\label{eq:Phi_G}\\
    f_\omega^{\Delta_\mathrm{g}}&=\frac{1}{\omega}\mathrm{e}^{-2\left(\frac{\Delta_\mathrm{g}}{\omega}\right)^2}\,,\label{eq:fomega}\\
    \beta_2&\approx\frac{\omega_\mathrm{g}^*}{\omega}+\pi\varepsilon_\mathrm{g,as}+\arctan\left(q\cot(\Theta_\mathrm{p})\right)+\frac{\pi}{4}\,,\label{eq:beta2}
\end{align}
where we replaced the unconstrained phase parameter $\delta$ introduced by these authors by $\pi\varepsilon_\mathrm{g,as}$, similarly as in Eq.\,(\ref{eq:delta}). Again, the phase contributed by a glitch in the buoyancy cavity is subtracted from the integer parameter in the eigenvalue condition. Like in Sect.\,\ref{ss:glitch_cavity}, the derivation of these terms relies on the asymptotic shape of the wave function around the glitch and the proportionality $k^2\propto\frac{N^2}{\omega^2}$ to isolate the glitch parameters. For instance, the summands $\frac{\omega_\mathrm{g}^*}{\omega}+\pi\varepsilon_\mathrm{g,as}$ in Eq.\,(\ref{eq:beta2}) originate from the approximation given in Eq.\,(\ref{eq:delta}). Therefore, it can only be expected to be valid once the spike has moved sufficiently far into the buoyancy cavity.

\section{Methods and data}\label{s:together}

In this section we summarize all the methods we used to test the various phase contributions.

\subsection{Stellar models}\label{ss:models}

\begin{figure}[t]
    \resizebox{\hsize}{!}{\includegraphics[width=\textwidth]{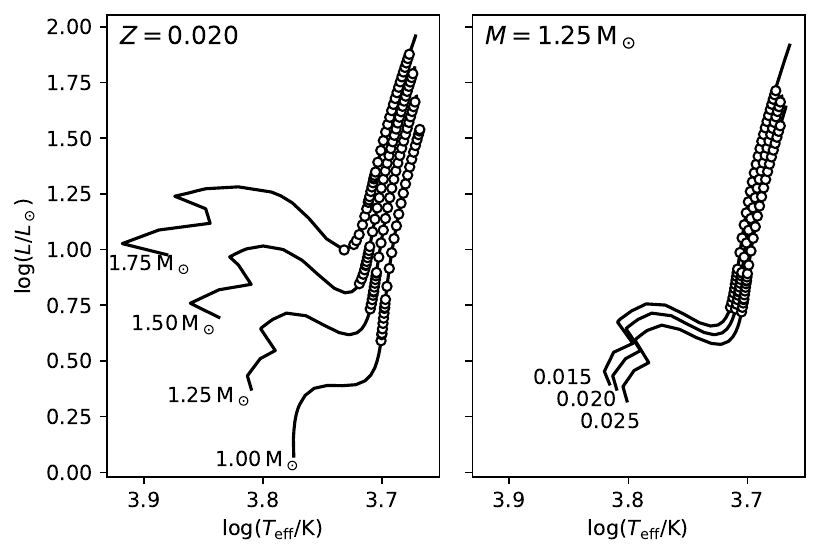}}
	\caption{Evolutionary tracks of stars with various masses $M$ at initial metallicity $Z=0.020$ (left panel, $M$ indicated in the figure) and various $Z$ at $M=1.25\,\mathrm{M}_\odot$ (right panel, $Z$ indicated in the figure). Circles mark the positions of the models for which g--mode frequency phases are shown in Sect.\,\ref{s:results}. The tracks are the same as the ones introduced by \citet{lit:vanLier2025}.}
	\label{fig:HRD_selections}
\end{figure}

The main ingredient for our comparisons are stellar models, for which we can calculate the different phase contributions from the stellar structure. We used version~r24.08.1 of the 1D stellar evolution code MESA \citep[`Modules for Experiments in Stellar Astrophysics';][]{lit:MESA1,lit:MESA2,lit:MESA3,lit:MESA4,lit:MESA5,lit:MESA6} to calculate canonical (i.e.~neglecting rotation, magnetic fields, overshooting and stellar winds) stellar models along evolutionary tracks with varying initial masses $M\in\{1.00,1.25,1.50,1.75\}\,\mathrm{M}_\odot$ (for $Z=0.020$) and metallicities $Z\in\{0.015,0.020,0.025\}$ (for $M=1.25\,\mathrm{M}_\odot$).\footnote{We used the same models as \citet{lit:vanLier2025}; the MESA inlists and \texttt{run\_star\_extras.f90} are available on Zenodo (doi: 10.5281/zenodo.15261756).} Hertzsprung-Russel diagrams with the resulting evolutionary tracks are shown in Fig.\,\ref{fig:HRD_selections}. The radial profile data of these models contain the characteristic frequencies and other ingredients needed to calculate all parameters for the phase contributions outlined in previous sections. Additionally, we got estimates for $\nu_\mathrm{max}$ from the scaling relation Eq.\,(\ref{eq:nu_max}), and for $\Delta\nu$ from the sound-crossing time through the star.

We further used version~7.2.1 of the stellar oscillation code GYRE \citep{lit:GYRE} to calculate the resonance frequencies of dipole modes within $[\nu_\mathrm{max}-3\Delta\nu,\nu_\mathrm{max}+3\Delta\nu]$ for all models, where we used the MESA estimates for $\nu_\mathrm{max}$ and $\Delta\nu$. This roughly corresponds to a typical observable frequency range. Since GYRE solves the oscillation equations numerically and does not rely on any asymptotic expansions, we consider these frequencies to accurately represent observed frequencies (given the structure of a star as given by the model profile).

\subsection{Evaluating theoretical prescriptions for $\varepsilon_\mathrm{g}$}\label{ss:epsg_theo}

For each of the models marked in Fig.\,\ref{fig:HRD_selections}, we calculated the g-mode offset as outlined in Sect.\,\ref{ss:epsg_numeric},
\begin{align}
    \varepsilon_\mathrm{g,num}=-\frac{1}{\pi}\int_{\tilde{r}}^{r_\mathrm{b}}\hspace{-0.5ex}k\cdot\left(2-\frac{1}{1-\left(\frac{2\pi\nu}{\hat{S}}\right)^2}-\frac{1}{1-\left(\frac{2\pi\nu}{\hat{N}}\right)^2}\right)\mathrm{d}r+\frac{\Psi}{\pi}\,, \label{eq:epsg_numeric}
\end{align}
where the integral still needs to be evaluated numerically as a Riemann sum. Since the integrand given by Eq.\,(\ref{eq:epsg_integrand}) goes to zero for $2\pi\nu\ll\hat{N},\hat{S}$ deep inside the buoyancy cavity, the region around $\tilde{r}$ contributes only little to the integral. Therefore the choice of this coordinate is expected not to affect the result significantly. To minimize the effect of the exact choice in $\tilde{r}$, we chose the model cell in which the integrand Eq.\,(\ref{eq:epsg_integrand}) had the minimum value as the lower integration limit. Numerical tests showed that the dependence decreases with evolution (which makes sense given that $\nu_\mathrm{max}$ decreases and therefore the assumption that $2\pi\nu_\mathrm{max}\ll\hat{N},\hat{S}$ is increasingly fulfilled), but even for the earliest models discussed in this work $\varepsilon_\mathrm{g}$ varies by less that $10^{-3}$ when changing $\tilde{r}$ on the order of $1\,\%$. This is below the numerical noise. For the upper integration bound, we used linear interpolation from the cells right below $r_\mathrm{b}$ to $r_\mathrm{b}$ to reduce the dependence on cell boundary positions. For the evanescent zone term $\Psi$ we implemented and compared all three versions discussed in Sect.\,\ref{ss:Psi}: the formula Eq.\,(\ref{eq:Psi_Takata}) as derived by \citet{lit:Takata2016a}, the wide--evanescent zone limit Eq.\,(\ref{eq:Psi_lim}), and the proposed modified $\hat{\Psi}$ from Eq.\,(\ref{eq:Psi_new}).

We also computed the g-mode offset analytically using the prescription introduced in Sect.\,\ref{ss:Pincon_model}. To this end, we fit a segmented model following Eq.\,(\ref{eq:N_Pincon}) to the modified Brunt-Väisälä frequency given by the stellar profile sufficiently above the hydrogen-burning shell and determined $\hat{N}_\mathrm{BCZ}$ from the fit. We then evaluated
\begin{align}
    \varepsilon_\mathrm{g,ana}=\left\{
	\begin{array}{ll}
		\varepsilon_\mathrm{g}^\mathrm{a}\,, & 2\pi\nu\geq\hat{N}_\mathrm{BCZ} \\
		\varepsilon_\mathrm{g}^\mathrm{b}\,, & 2\pi\nu<\hat{N}_\mathrm{BCZ}
	\end{array}
    \label{eq:epsg_analytic}
	\right.\,,
\end{align}
with $\varepsilon_\mathrm{g}^\mathrm{a}$ and $\varepsilon_\mathrm{g}^\mathrm{b}$ given by Eqs.\,(\ref{eq:epsg_Pincon_a}) and~(\ref{eq:epsg_Pincon_b}), respectively.

We calculated both $\varepsilon_\mathrm{g,num}$ and $\varepsilon_\mathrm{g,ana}$ for $\nu=\nu_\mathrm{max}$ as well as for all dipole mode frequencies computed by GYRE. This way we were able to estimate a range of values expected across the observable range of modes.

\subsection{Fitting for the phase}\label{ss:fitting}

To simulate the asymptotic parameters as they would be derived from seismic observations and link them to the calculation from the stellar models, we fit the asymptotic periods (Eq.\,\ref{eq:Pi_as_mixed}) to the inverse of the models' observed frequencies. The full fitting procedure is outlined in \citet{lit:vanLier2025}, we summarize it again in Appendix~\ref{ass:noglitch_fitting}. Typically, the phase $\Phi$ obtained by such a procedure is identified with the gravity mode offset $\varepsilon_\mathrm{g}+\frac{1}{2}$, even though it should contain any phase contribution (except for the phase contributed by the coupling to p modes, which is explicitly isolated in $\Pi_\mathrm{as}$), including for example that by glitches. With the derivations presented in this work, we can write the phase as a sum of various contributions,
\begin{align}
    \Phi=\varepsilon_\mathrm{g}+\frac{1}{2}-\Phi_\mathrm{cavity}+\Phi_\mathrm{EZ}\,,
\end{align}
where $\Phi_\mathrm{cavity}$ is a placeholder for a choice of $\Phi_\delta$ or $\Phi_\mathrm{G}$. Whenever the spike in the Brunt-Väisälä frequency did not lie in the evanescent zone of a model, we chose $\Phi_\mathrm{EZ}\equiv 0$; when it was not in the buoyancy cavity, we set $\Phi_\mathrm{cavity}\equiv 0$. To test whether any other contributions to this phase are missing, we isolated $\varepsilon_\mathrm{g}$ and compared it to the theoretically predicted values.

When both glitch contributions to the phase were zero, we could identify the g-mode offset from an asymptotic fit as:
\begin{align}
    \varepsilon_\mathrm{g,as}=\Phi-\frac{1}{2}\quad\text{without a glitch.}
\label{eq:epsg_as_noglitch}
\end{align}
The calculations with a glitch in the evanescent zone were similarly straightforward. For $\hat{N}\ll2\pi\nu\ll\hat{S}$, which would be given deep inside a wide evanescent zone, Eq.\,(\ref{eq:dispersion_relation}) can be expanded to $\kappa\approx\frac{\sqrt{2}}{r}$. While numeric tests showed that this approximation for the decay length deviates significantly from the full expression for our stellar models, we still observed that the frequency dependence of $\kappa$ was small, as suggested by this limit. We therefore assumed it to be sufficiently small so we could calculate $\Phi_\mathrm{EZ}$ (Eq.\,\ref{eq:Phi_EZ}) just at $\nu_\mathrm{max}$ to obtain an estimate suitable across the full range of fit frequencies. Since we assume $\Phi_\mathrm{EZ}$ to be frequency-independent, it becomes fully degenerate with $\varepsilon_\mathrm{g,as}$. Therefore we fixed its value for each model according to Eq.\,(\ref{eq:Phi_EZ}) using the stellar profile and subtracted it from the phase $\Phi-\frac{1}{2}$ obtained in the fit to get an estimate for $\varepsilon_\mathrm{g,as}$:
\begin{align}
    \varepsilon_\mathrm{g,as}=\Phi-\frac{1}{2}-\Phi_\mathrm{EZ}\quad\text{with a glitch in the evanescent zone.}
\label{eq:epsg_as_EZglitch}
\end{align}
$\Phi_{\delta}$ and $\Phi_\mathrm{G}$ on the other hand are explicitly frequency-dependent, even though the parameters $\mathfrak{C}$ and $\omega_\mathrm{g}^*$ (or $C_\mathrm{G}$, $\Delta_\mathrm{g}$, and $\omega_\mathrm{g}^*$) are only defined by the profile of $N^2$ and hence independent of the oscillation frequency.\footnote{We note that $\omega_\mathrm{g}^*$ implicitly depends on frequency via the definition of the turning point. However, since the spike forms at the lowest extent of the convective envelope, which has not receded much in the models we study, we can reasonably assume that $r_\mathrm{b}\approx r_\mathrm{BCZ}$ (see also Fig.\,\ref{fig:propagation_diagram}), and hence $\omega_\mathrm{g}^*$ is only weakly frequency-depend.}  While \citet{lit:Pincon2019} argued that $\Phi_\mathrm{cavity}$ should be small compared to observational uncertainties in the phase offset, we want to study whether this frequency dependence and resulting different shape of the eigenfrequency pattern can still affect the convergence of the fit. Therefore, we included the explicit frequency dependence into the asymptotic period formula (Eq.\,\ref{eq:Pi_as_mixed}). The specifics of the new fitting are described in Appendix~\ref{ass:glitch_fitting}.

\subsection{Observational data}\label{ss:observations}

The asteroseismic catalog published by \citet{lit:Mosser2018} contains measurements of $\varepsilon_\mathrm{g}$ in the sense $\varepsilon_\mathrm{g}=\Phi-\frac{1}{2}$ for more than 150~RGB stars observed by the \textit{Kepler} mission. We use these measurements to cross-check our results and verify whether stellar models can accurately reproduce the relevant properties of real stars. The dataset also contains estimates for $\nu_\mathrm{max}$, which we use as an indicator of evolution along which we can compare models and observations.

In order to identify any potential mass trends, we used the KIC~number to cross-match the sources with \makebox{APOKASC-2} \citep{lit:APOKASC2}, which contains mass estimates for many solar-like oscillators in the \textit{Kepler} field. We found 114 RGB stars for which we could obtain both a mass and a g-mode offset from the combination of these two catalogs.

\section{Results}\label{s:results}

\begin{figure}[t]
    \resizebox{\hsize}{!}{\includegraphics[width=\textwidth]{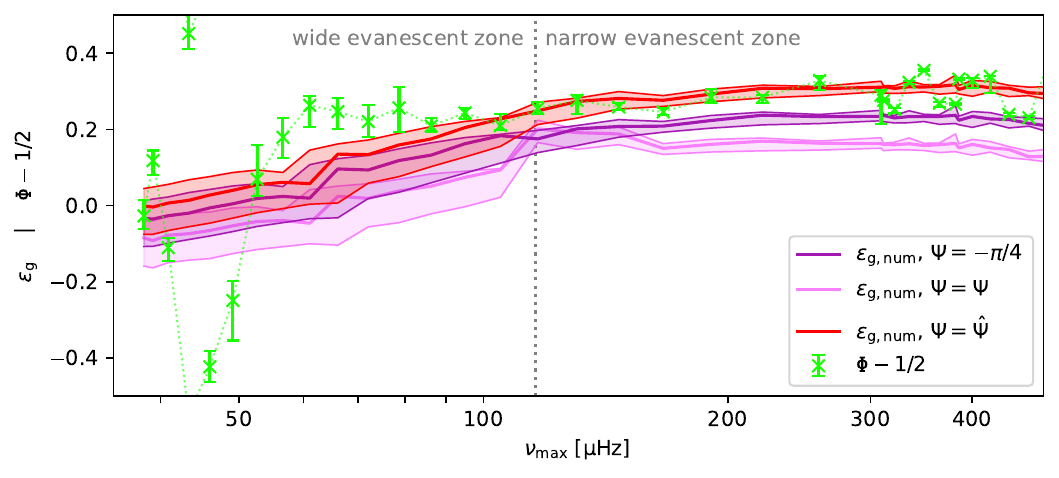}}
	\caption{Numerically evaluated g-mode offset compared to the asymptotic phase shift `observed' from frequencies of models marked in Fig.\,\ref{fig:HRD_selections} along the $M=1.25\,\mathrm{M}_\odot,Z=0.020$ track, as a function of $\nu_\mathrm{max}$. Stars evolve from right to left. Green crosses with error bars show results of the frequency fitting with uncertainties as defined in Appendix~\ref{ass:noglitch_fitting}. Thick purple, pink and red lines show $\varepsilon_\mathrm{g,num}(\nu_\mathrm{max})$ with $\Psi$ in the wide--evanescent zone limit, as derived by \citet{lit:Takata2016a}, and as defined in Eq.\,(\ref{eq:Psi_new}), respectively; corresponding shaded areas show the values obtained across the frequency range used in the fits. The vertical dotted line shows the transition from the regime of a narrow to a wide evanescent zone as defined by \citet{lit:vanLier2025}.}
	\label{fig:Psi_comparison}
\end{figure}

In Fig.\,\ref{fig:Psi_comparison}, we show a comparison between the g-mode offset calculated numerically from the stellar model profiles as outlined in Sect.\,\ref{ss:epsg_numeric} to the asymptotic fit for the phase $\Phi-\frac{1}{2}$. The different formulations for the evanescent-zone term $\Psi$ presented in Sect.\,\ref{ss:Psi} differ most strongly on the early RGB (corresponding to larger $\nu_\mathrm{max}$) since the coupling is strongest in this regime. The comparison to $\Phi-\frac{1}{2}$ shows that the commonly used weak-coupling limit (Eq.\,\ref{eq:Psi_lim}) is indeed not fully applicable there. For our numerical evaluation, $\hat{\Psi}$ is the most suitable prescription. Therefore, whenever we plot $\varepsilon_\mathrm{g,num}$ as given by Eq.\,(\ref{eq:epsg_numeric}) in the following, we use $\Psi=\hat{\Psi}$ (Eq.\,\ref{eq:Psi_new}). For more evolved models, $\varepsilon_\mathrm{g,num}$ converges to agree between all three expressions. However, for $\nu_\mathrm{max}\la100\,\text{\textmu Hz}$ the asymptotic phase no longer follows the trend predicted by our numerical calculations.

\begin{figure}[t]
    \resizebox{\hsize}{!}{\includegraphics[width=\textwidth]{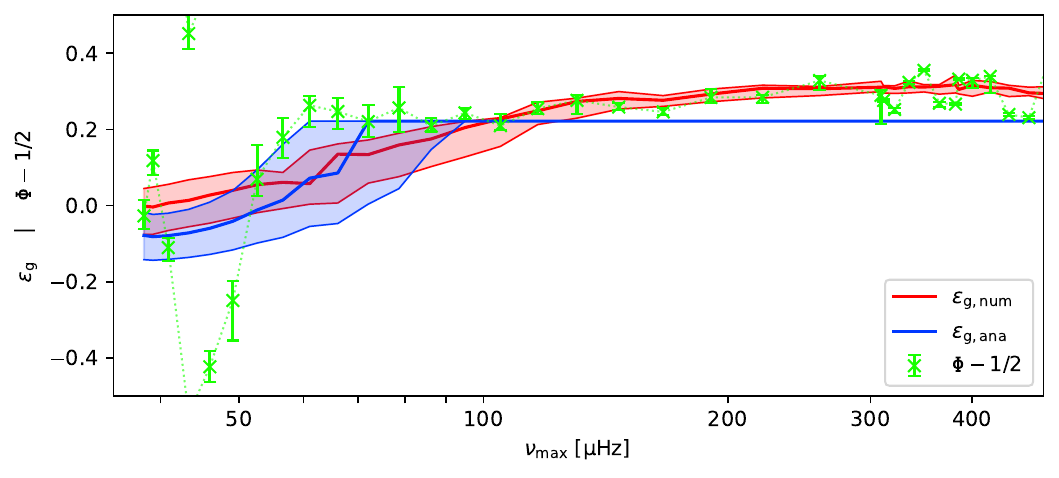}}
	\caption{Analytically (blue) and numerically (red) evaluated g-mode offset compared to the asymptotic phase shift `observed' from the frequencies of models marked in Fig.\,\ref{fig:HRD_selections} along the $M=1.25\,\mathrm{M}_\odot,Z=0.020$ track, as a function of $\nu_\mathrm{max}$. Stars evolve from right to left. See Fig.\,\ref{fig:Psi_comparison} for the meaning of different symbols and lines.}
	\label{fig:epsg_Pincon}
\end{figure}

The result of estimating the g-mode offset from an analytical model as described in Sect.\,\ref{ss:Pincon_model} is shown in Fig.\,\ref{fig:epsg_Pincon}. On the early RGB ($\nu_\mathrm{max}\ga100\,\text{\textmu Hz}$), $\varepsilon_\mathrm{g,ana}$ as given by Eq.\,(\ref{eq:epsg_analytic}) correctly reproduces the trend that the g--mode frequency phase is roughly constant, as it is also visible in both $\Phi-\frac{1}{2}$ and $\varepsilon_\mathrm{g,num}$. However, the application of the weak-coupling limit for the evanescent zone term means that the value is systematically underestimated. On the late RGB ($\nu_\mathrm{max}\la100\,\text{\textmu Hz}$), the analytical and numerical calculation of the g-mode offset give consistent values within the range of observed frequencies. The analytical model can therefore be used to estimate $\varepsilon_\mathrm{g}$ for a given model in this regime. However, the morphology of the time evolution is significantly different: While $\varepsilon_\mathrm{g,num}$ decreases smoothly with evolution, $\varepsilon_\mathrm{g,ana}$ stalls at the constant value from the early RGB for longer, before showing a kink and steep decrease. The latter trend appears to agree with the evolution of $\Phi-\frac{1}{2}$ better, however, neither theoretical prescription for the g-mode offset reproduces the extreme values the observed phase takes in this regime.

\begin{figure*}[t]
    \centering
    \includegraphics[width=17cm]{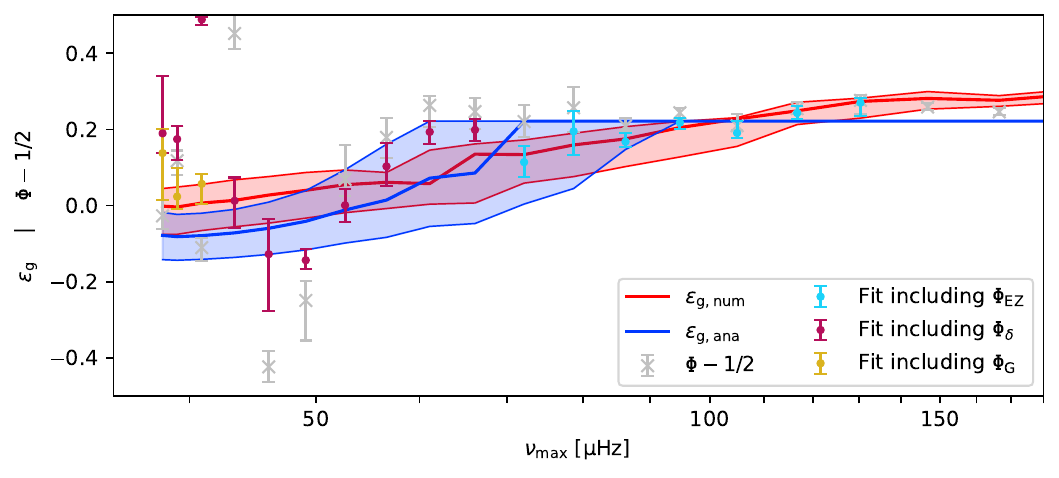}
     \caption{Asymptotic g-mode offsets from fits including glitch phase contributions compared to $\varepsilon_\mathrm{g,num}$ and $\varepsilon_\mathrm{g,ana}$, as a function of $\nu_\mathrm{max}$. Stars evolve from right to left. Light blue markers show results including the phase as introduced in Sect.\,\ref{ss:glitch_EZ} for a glitch in the evanescent zone; for a glitch in the buoyancy cavity, the prescriptions used are the ones described in Sects.\,\ref{ss:glitch_cavity} (purple) and~\ref{ss:glitch_GaussCunha} (ocher), respectively. For reference, the observed phase as in Fig.\,\ref{fig:Psi_comparison} is shown in gray. Uncertainties on $\varepsilon_\mathrm{g,as}$ as defined in Appendix~\ref{ass:glitch_fitting}.}
     \label{fig:glitchfit_results}
 \end{figure*}

Figure \ref{fig:glitchfit_results} shows the effect of including the glitch effects to calculate $\varepsilon_\mathrm{g,as}$. Since we used Eq.\,(\ref{eq:epsg_as_noglitch}) on the early RGB, only a subset of models is shown compared to Fig.\,\ref{fig:epsg_Pincon}. We observe that including the glitches into the fitting significantly improves the agreement between $\varepsilon_\mathrm{g,as}$ and $\varepsilon_\mathrm{g,num}$. Subtracting $\Phi_\mathrm{EZ}$ causes the phase to drop below the original constant value earlier, in good agreement with the evolution of the numerically calculated g-mode offset. Once the spike in the Brunt-Väisälä frequency has moved into the buoyancy cavity, the glitch model using $\Phi_{\delta}$ \citep[see also Sect.\,\ref{ss:glitch_cavity}]{lit:Cunha2015} significantly improves the agreement between fit and numerical calculation, however there are still discrepancies. The agreement for the most evolved models with $\nu_\mathrm{max}\la40\,\text{\textmu Hz}$ can be improved using the Gaussian glitch prescription $\Phi_\mathrm{G}$ \citep[see also Sect.\,\ref{ss:glitch_GaussCunha}]{lit:Cunha2019}, as demonstrated by the ocher points in Fig.\,\ref{fig:glitchfit_results}. We discuss these findings in the following section.

\section{Discussion}\label{s:discussion}

From the theoretical perspective on the g-mode offset, the results presented in Fig.\,\ref{fig:Psi_comparison} clearly show that the weak-coupling limit $\Psi\rightarrow-\frac{\pi}{4}$ for the evanescent-zone term is not applicable on the early RGB. This is to be expected since the evanescent zone is relatively narrow during this evolutionary stage \citep{lit:vanRossem2024}. The difference between this limit and the numerical evaluation of $\Psi$ given by Eq.\,(\ref{eq:Psi_new}) ($\sim0.06$) is significant compared to some of the observational uncertainties achieved for $\Phi-\frac{1}{2}$ by \citet{lit:Mosser2018}; with future advances in observational techniques the significance is likely to increase. It will therefore become necessary to correctly account for the evanescent-zone contribution to the g--mode frequency phase. Our model computations suggest that the formulation we derived in Eq.\,(\ref{eq:Psi_new}) is promising to match observationally obtained phases. The fact that the three prescriptions for $\Psi$ shown in Fig.\,\ref{fig:Psi_comparison} converge as evolution progresses can be explained by the fact that the evanescent zone widens \citep{lit:vanRossem2024} and therefore the phase contribution by the coupling to the p-mode cavity decreases.

The results presented in Fig.\,\ref{fig:epsg_Pincon} suggest that the analytical model might reproduce the morphology of the time evolution of the g--mode frequency phase better. This is in line with the findings by \citet{lit:Pincon2019}, who compared similarly obtained $\varepsilon_\mathrm{g,ana}$ to an ensemble of observed g-mode offsets. However, the results shown in Fig.\,\ref{fig:glitchfit_results} demonstrate that this leads to a misinterpretation of the observed phases. While \citet{lit:Pincon2019} concluded that the kink as the phase starts to drop is inherent to the g-mode offset produced by the behavior of the wave function at the outer boundary of the (smooth) cavity (Eq.\,\ref{eq:epsg_Takata}), we were able to show that this feature is more likely linked to the spike in the Brunt-Väisälä frequency acting as a glitch. We were further able to support this hypothesis by calculating an additional set of stellar models where we artificially removed the spike. For these models, the phase $\Phi-\frac{1}{2}$ did not show a kink and closely followed the respective $\varepsilon_\mathrm{g,num}$, while $\varepsilon_\mathrm{g,ana}$ still had a similar shape to the models including the spike (see Fig.\,\ref{fig:epsg_nomudisc}). 

While the glitch is in the evanescent zone, our calculations support that subtracting the phase term derived in Sect.\,\ref{ss:glitch_EZ} evaluated at $\nu_\mathrm{max}$ from the glitch-free observed phase $\Phi-\frac{1}{2}$ (Eq.\,\ref{eq:epsg_as_EZglitch}), is suitable to isolate a g-mode offset consistent with $\varepsilon_\mathrm{g,num}$. Since $\Phi_\mathrm{EZ}$ is straightforward to calculate from stellar model profiles, it enables to predict observed g--mode frequency phases as $\Phi-\frac{1}{2}=\varepsilon_\mathrm{g,num}+\Phi_\mathrm{EZ}$. We note that we observe this good agreement in our models for which we can calculate and fit all possible dipole modes. In real observations, only modes with sufficient power in the p-mode cavity are detectable. These modes are also those for which the coupling-induced phase is the largest, and thus the assumption that we can derive $\Phi_\mathrm{EZ}$ for the isolated g-mode component is potentially least appropriate for them. We did test, however, how the fit for $\Phi-\frac{1}{2}$ is affected by restricting the mode sample to more p-dominated modes and did not find a significant difference. This suggests that the treatment of the glitch in the evanescent zone presented in this work can also suffice to interpret observed g-mode offsets.

For more evolved models, when the spike has moved into the cavity, we generally observe improved agreement between asymptotic fits and theory when introducing any version of $\Phi_\mathrm{cavity}$. While the prescription using $\Phi_\delta$ (introduced in Sect.\,\ref{ss:glitch_cavity}) does not allow to isolate a g-mode offset fully consistent with $\varepsilon_\mathrm{g,num}$, for all but the three most evolved models it brings $\varepsilon_\mathrm{g,as}$ closer to the value calculated from the stellar profiles (cf. the purple points versus the gray crosses in Fig.\,\ref{fig:glitchfit_results}). As discussed in Sect.\,\ref{ss:glitch_cavity}, the derivation of this prescription relies on three assumptions that are not completely fulfilled in some of the models, which could explain the remaining discrepancies.

Firstly, the prescription assumes that the wave function takes its asymptotic form either side of the glitch. This is only the case when the spike has moved more than approximately $k^{-1}$ into the cavity, which would correspond to $\int_{r^*}^{r_\mathrm{b}}k_0\mathrm{d}r\sim1$ and therefore does not hold for the earliest models with a glitch in the oscillation cavity (cf.\,bottom panel of Fig.\,\ref{fig:glitch_parameters}).

Secondly, we adopted the expansion of the asymptotic wave vector in the limit $\omega^2\ll N^2,S^2$ valid deep inside the buoyancy cavity. This approximation enters in two places. On the one hand, it is used to isolate the frequency dependence of the glitch amplitude in Eq.\,(\ref{eq:Phi_Cunha}) such that the amplitude can be written as $\frac{\mathfrak{C}}{\omega}$. The top panel of Fig.\,\ref{fig:glitch_parameters} shows that this approximation agrees with the amplitudes estimated from the shape of the spike in the full asymptotic wave vector (cf.\,Eq.\,\ref{eq:dispersion_relation}) at $\nu_\mathrm{max}$ within $\lesssim10\,\%$ for all models. On the other hand, the approximation in Eq.\,(\ref{eq:delta}) relies on the assumption that the limit is met everywhere between $r^*$ and $r_\mathrm{b}$ except for a region very near the turning point, and that the correction for both the deviation of $\frac{\sqrt{2}N}{\omega r}$ from the full asymptotic wave vector in that regime as well as the non-asymptotic behavior of the wave function near $r_\mathrm{b}$ can be absorbed into a constant phase given by the g-mode offset. The bottom panel of Fig.\,\ref{fig:glitch_parameters} shows that this assumption is increasingly appropriate as the star evolves. We note that the difference shown in the inset would never go to zero, because we compare the first two terms in Eq.\,(\ref{eq:delta}) to an integral of the asymptotic wave vector which does not include non-asymptotic corrections. Again, this approximation results in an isolation of the frequency dependence of $\phi$, such that all free glitch parameters $\mathfrak{C}$, $\omega_\mathrm{g}^*$ (and $\varepsilon_\mathrm{g}$) are frequency-independent, which enabled us to explore the parameter space in our fitting. However, the remaining discrepancies between $\varepsilon_\mathrm{g,as}$ and $\varepsilon_\mathrm{g,num}$ in the regime $40\,\text{\textmu Hz}\la\nu_\mathrm{max}\la 70\,\text{\textmu Hz}$ can likely be explained by violations of the assumptions discussed so far: the wave function not taking its asymptotic form around $r^*$, the frequency dependence of the glitch amplitude being estimated incorrectly by taking the limit $\omega^2\ll N^2,S^2$, and the phase shift $\pi\varepsilon_\mathrm{g,as}$ not adequately correcting for the near--turning point behavior. We therefore see the results discussed here more as an indication that the observed phases need to be interpreted accounting for a glitch, while reliably isolating the pure g-mode offset from observations will require a different approach.

For the three most evolved models, the third assumption becomes relevant. As the star evolves, diffusive mixing smooths the chemical discontinuity to a finite width and the spike moves deeper into the buoyancy cavity where the local wavelength is shorter, causing the glitch to widen over time. Therefore, modeling the glitch by a Dirac-delta function is not suitable for the most evolved models, which likely explains the significant disagreement between $\varepsilon_\mathrm{g,as}$ accounting for $\Phi_\delta$ and $\varepsilon_\mathrm{g,num}$ for $\nu_\mathrm{max}\la40\,\text{\textmu Hz}$. Indeed, we were able to show that using the finite--width glitch prescription $\Phi_\mathrm{G}$ as introduced by \citet{lit:Cunha2019} mitigates these differences. For even more evolved models, the width of the spike becomes so large relative to the local wavelength that it no longer acts as a glitch.

We note that including the phase contributed by a glitch in the cavity can shift $\varepsilon_\mathrm{g,as}$ to both higher and lower values compared to $\Phi-\frac{1}{2}$ depending on the model (cf.~Fig.\,\ref{fig:glitchfit_results}), despite $\Phi_\mathrm{cavity}\geq0$ for both prescriptions. This demonstrates that including it to the fitting does not only induce an absolute phase shift, but that also the changed frequency pattern affects the convergence of the fit. Therefore, our results do not contradict the estimate by \citet{lit:Pincon2019}, who argued that the phase contributed by glitches should be small compared to $\varepsilon_\mathrm{g}$, while differences $\left|(\Phi-\frac{1}{2})-\varepsilon_\mathrm{g,num}\right|\gg\Phi_\mathrm{cavity}$ can still be explained by the presence of a glitch.

Both Eqs.\,(\ref{eq:epsg_numeric}) and~(\ref{eq:epsg_Pincon_b}) for the numerical g-mode offset and $\varepsilon_\mathrm{g,ana}$ in case b clearly depend on frequency. \citet{lit:Pincon2019} estimated that this dependence is small compared to observational uncertainties for the g--mode frequency phase. The representations of the theoretical g-mode offset in Figs.\,\ref{fig:Psi_comparison}--\ref{fig:glitchfit_results} show the frequency dependence of $\varepsilon_\mathrm{g,num}$ by the width of the shaded area around $\varepsilon_\mathrm{g,num}(\nu_\mathrm{max})$, which represents the range of values obtained for frequencies in the sample used for the asymptotic fit. The frequency dependence is indeed quite small on the early RGB, but becomes stronger with evolution until the range of g-mode offsets across $[\nu_\mathrm{max}-3\Delta\nu,\nu_\mathrm{max}+3\Delta\nu]$ (a typical observable frequency range) becomes $\sim 0.1$ in width. This evolution can be explained using the arguments for the analytical model in Sect.\,\ref{ss:Pincon_model}: The early RGB corresponds to case a, which is expected to have a weak frequency dependence, while in case b the frequency affects the g-mode offset. A range of $\sim0.1$ is comparable to the precision achieved by \citet{lit:Mosser2018} for some stars. With both observations and asteroseismic analysis advancing in the future, the frequency dependence of $\varepsilon_\mathrm{g}$ will therefore likely become important to correctly interpret $\Phi-\frac{1}{2}$. The fact that we also neglected it in our fitting algorithm to be comparable to current observational techniques is another potential explanation for the remaining discrepancies in Fig.\,\ref{fig:glitchfit_results}.

\subsection{Mass and metallicity dependence}\label{ss:MZ_dependence}

We tested the influence of mass and metallicity on the results discussed above, using the models marked along all the evolutionary tracks in Fig.\,\ref{fig:HRD_selections}. Since the results are qualitatively similar to the ones for $M=1.25\,\mathrm{M}_\odot, Z=0.020$ presented before, we merely summarize them in Fig.\,\ref{fig:epsg_MZ}.

Across all masses and metallicities, the fit values $\Phi-\frac{1}{2}$ cluster around a roughly constant value well reproduced by $\varepsilon_\mathrm{g,num}$ on the early RGB. As the stars evolve, $\Phi-\frac{1}{2}$ shows a kink into a steep drop at some point. The corresponding value of $\nu_\mathrm{max}$ is also the main difference between the evolutionary tracks, with the drop occurring at lower $\nu_\mathrm{max}$ for higher masses. Since this feature is related to the spike in the Brunt-Väisälä frequency moving from the evanescent zone into the g-mode cavity, this is consistent with \citet{lit:vanLier2025}, who found that the spike appears in the evanescent zone at lower $\nu_\mathrm{max}$ for higher masses.

Including the glitch into the fitting improves the agreement between $\varepsilon_\mathrm{g,as}$ and $\varepsilon_\mathrm{g,num}$, removing the kink as a function of $\nu_\mathrm{max}$ and making the fit phase values less extreme. The shortcomings of $\Phi_{\delta}$ are also similar for all models. Along all tracks, $\Phi_\mathrm{EZ}$ as derived in Sect.\,\ref{ss:glitch_EZ} gives good results with one exception: the most evolved model with $\Phi_\mathrm{EZ}\not=0$ along the $M=1.25\,\mathrm{M}_\odot,Z=0.015$ track (cf.~top right panel of Fig.\,\ref{fig:epsg_MZ}). In this particular model, the glitch in the evanescent zone happens to be located very close to the cavity boundary. Therefore, the approximation that the parameters $\mathfrak{A}$ and $\mathfrak{d}^*$ are frequency-independent likely breaks down and subtracting a constant $\Phi_\mathrm{EZ}$ was no longer a suitable method to isolate $\varepsilon_\mathrm{g,as}$.

\subsection{Comparison to observations}\label{ss:discuss_observations}

\begin{figure}[t]
    \resizebox{\hsize}{!}{\includegraphics[width=\textwidth]{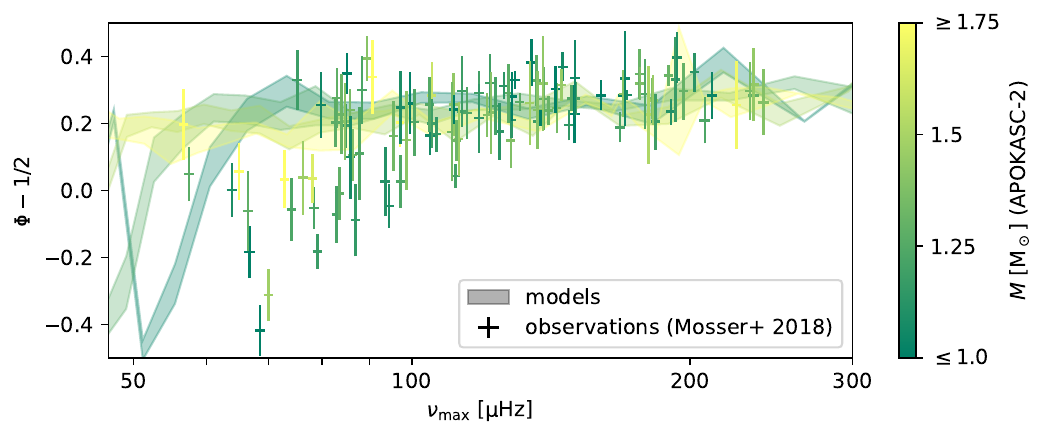}}
	\caption{Comparison of observed phase $\Phi-\frac{1}{2}$ between observations by \citet{lit:Mosser2018} and our models marked along the evolutionary tracks with $Z=0.020,M\in\{1.00,1.25,1.50,1.75\}\,\mathrm{M}_\odot$ (cf.~left panel of Fig.\,\ref{fig:HRD_selections}) as a function of $\nu_\mathrm{max}$, color-coded by stellar mass. Stars evolve from right to left. The displayed range of $\Phi-\frac{1}{2}$ for the models relates to the uncertainties as discussed in Appendix~\ref{ass:noglitch_fitting}. Stellar masses for observed stars are taken from the \makebox{APOKASC-2} catalog, typical uncertainties are $\sim0.08\,\mathrm{M}_\odot$ \citep{lit:APOKASC2}.}
	\label{fig:epsg_observations}
\end{figure}

Having empirically shown that the g-mode offset and the glitch produced by the chemical discontinuity left behind by the first dredge-up are the dominant contributions to the g--mode frequency phase in our stellar models, we aimed to verify whether our models are able to reproduce the phase observed for real stars, and therefore whether these two contributions also suffice to interpret observations. To this end, in Fig.\,\ref{fig:epsg_observations} we compare the fit phases $\Phi-\frac{1}{2}$ from our models with various masses to the observations mentioned in Sect.\,\ref{ss:observations}.

Neither the observations nor the observed phases from our models show a mass dependence for $\nu_\mathrm{max}\ga100\,\text{\textmu Hz}$ -- they all cluster around a constant value. In this regime on the early RGB, observations and models agree well. The mean of observed phases in the range $\nu_\mathrm{max}\in[100,300]\,\text{\textmu Hz}$ is given by
\begin{align}
	\left\langle\Phi-\frac{1}{2}\right\rangle_\mathrm{obs.}=0.269\pm0.011\,,\quad
	\left\langle\Phi-\frac{1}{2}\right\rangle_\mathrm{mod.}=0.251\pm0.007\,,
	\label{eq:constant_Phi_mean}
\end{align}
respectively. These values agree within $1.4\,\sigma$.

As discussed above, the position of the drop in phase values predicted by our models is mass-dependent. The observational data at sufficiently low $\nu_\mathrm{max}$ are too sparse to conclude whether a similar mass dependence is present in the ensemble of real stars. However, the decrease of $\Phi-\frac{1}{2}$ clearly sets in significantly earlier for the observed stars than for our models. Following the arguments by \citet{lit:Pincon2019}, this would suggest that $\hat{N}_\mathrm{BCZ}$ is predicted incorrectly by the models. Since we showed that the morphology of the phase as a function of $\nu_\mathrm{max}$ is closely related to the spike of the Brunt-Väisälä frequency, however, we can instead interpret this observation as an indication that the spike becomes relevant as a glitch earlier (at higher $\nu_\mathrm{max}$) than our models predict. This corresponds to the convective envelope reaching further down into the star during the first dredge-up.

\begin{figure}[t]
    \resizebox{\hsize}{!}{\includegraphics[width=\textwidth]{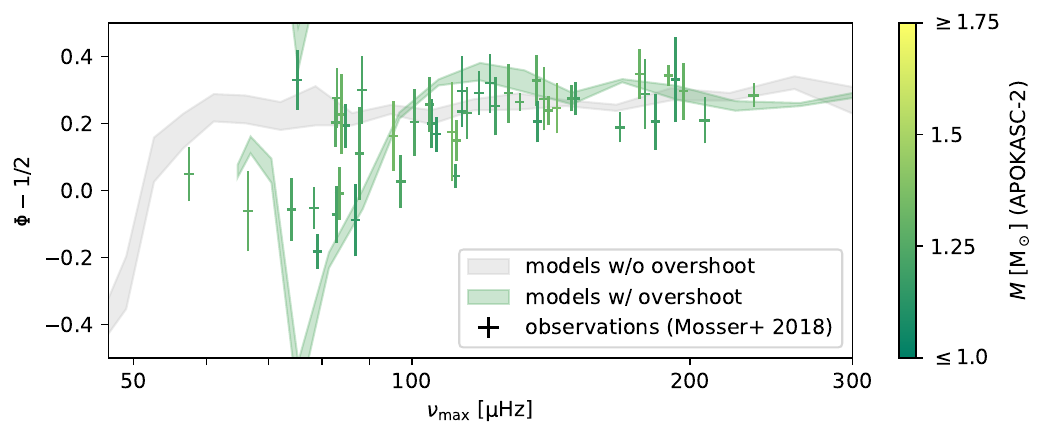}}
	\caption{Observed g-mode phase $\Phi-\frac{1}{2}$ affected by overshooting. Markers with error bars show the sub-sample of observations by \citet{lit:Mosser2018} with $1.15\,\mathrm{M}_\odot\leq M\leq1.35\,\mathrm{M}_\odot$ \citep{lit:APOKASC2}, the gray shaded area shows the phases from the fit to the original $M=1.25\,\mathrm{M}_\odot$ models (cf.~Figs~\ref{fig:epsg_Pincon},~\ref{fig:epsg_observations}) and the green shaded area shows the fit to models including overshooting.}
	\label{fig:epsg_overshooting}
\end{figure}

One possible modification to the model physics to achieve this shift of the spike is to include convective overshooting, which extends the chemical mixing beyond the mathematical boundary of the convective envelope. We calculated an evolutionary track with $M=1.25\,\mathrm{M}_\odot,Z=0.020$ using the same settings as described in Sect.\,\ref{ss:models}, now including exponential overshooting with a decay constant of 0.04 pressure-scale heights, starting from 0.01 pressure-scale heights above the convective boundary to test this possibility. Figure~\ref{fig:epsg_overshooting} shows that, while we did not yet fully reproduce the evolution of the g--mode frequency phase of the ensemble observations with our modified stellar models, we were able to increase agreement by shifting the spike in the Brunt-Väisälä frequency in this way. Therefore, we conclude that the two dominant phase contributions we identified for the stellar models are also sufficient to describe observed g--mode frequency phases in ensembles of stars. We note that \citet{lit:Villate2026} showed that for individual stars with strong core magnetic fields, this magnetic field also impacts the observed phase $\Phi$. Previously, \citet{lit:Bugnet2022} had shown the effect of core magnetic fields on the period spacing $\Delta\Pi$.

With the glitch position linked to the evolution of convective boundaries in stellar models as demonstrated above, and the shape of the glitch clearly dependent on the chemical mixing, the g-mode phase even becomes a potentially interesting observable to constrain physical processes in stars and their implementation in stellar evolution codes, especially when the sample of reliable observations grows in the future. The morphology of $\Phi-\frac{1}{2}$ as a function of $\nu_\mathrm{max}$ could be parametrized in terms of different mixing schemes and parameters, extending on our ad-hoc demonstration in Fig.\,\ref{fig:epsg_overshooting}.

\section{Conclusions}\label{s:conclusions}

By comparing theoretical contributions to the phase offset of g-mode eigenfrequencies to the phases observed from models, we constrained the interpretation of the g-mode offset. Firstly, we showed that the common wide--evanescent zone expansion $\Psi\rightarrow-\frac{\pi}{4}$ for the evanescent-zone contribution to $\varepsilon_\mathrm{g}$ is not applicable on the early RGB. However, it seems to be valid as a limit as the evanescent zone widens with evolution. We get results most consistent with the observed phase when using Eq.\,(\ref{eq:Psi_new}) to calculate the evanescent-zone term, rather than the term originally derived by \citet{lit:Takata2016a}, given in Eq.\,(\ref{eq:Psi_Takata}). The assumption we made when deriving the modified version $\hat{\Psi}$ will also affect the wave function in the evanescent zone, further studies would be required to investigate how this affects other results derived by \citet{lit:Takata2016a}, including the coupling strength in the strong-coupling limit.

Secondly, the frequency dependence of $\varepsilon_\mathrm{g}$ as calculated from our models is significant on the late RGB. It becomes comparable with observational uncertainties and therefore more elaborate fitting procedures might need to take it into account.

Thirdly, the analytical model of the buoyancy cavity introduced by \citet{lit:Pincon2019} is suitable to estimate values of the g-mode offset within observational uncertainties on the late RGB. However, it predicts a trend in the evolution of $\varepsilon_\mathrm{g}$ with $\nu_\mathrm{max}$ that is not consistent with the numerical evaluation of the full expression for the g-mode offset. The trend in the analytically estimated g-mode offset partly reproduces the effect of the Brunt-Väisälä spike acting as a glitch in the evanescent zone and therefore leads to a misinterpretation of the observed phases. While \citet{lit:Pincon2019} suggest that the kink in $\Phi-\frac{1}{2}$ at the transition from $r_\mathrm{b}$ being frequency-dependent to $r_\mathrm{b}\equiv r_\mathrm{BCZ}$ is inherent to Eq.\,(\ref{eq:epsg_Takata}), we were able to demonstrate that it is instead produced by the phase contribution from the glitch.

Lastly, we were able to show that the Brunt-Väisälä spike contributes significantly to the phase offset of g-mode frequencies. When it acts as a glitch in the buoyancy cavity, not only the additive phase $\Phi_\mathrm{cavity}$ but also the changed pattern of the frequency spectrum lead to a shift in the observed $\Phi-\frac{1}{2}$ relative to the g-mode offset $\varepsilon_\mathrm{g}$ given by the (smooth) shape of the buoyancy cavity. It becomes clear that in this way the g-mode phase is an observable to investigate both, the shape of the Brunt-Väisälä spike and the treatment of convective boundaries in stellar models by the remaining discrepancies in a model-to-model and model-to-observation comparison, respectively.

\begin{acknowledgements}
      We thank Margarida Cunha for her kind and constructive referee reports and insightful suggestions which significantly improved the rigor of the presented work. The research leading to the presented results has received funding from the European Research Council Consolidator Grant DipolarSound (grant agreement No. 101000296).
\end{acknowledgements}

\bibliography{aa58519-25_Literature}

\begin{appendix}

\section{Asymptotic fitting of mixed modes}\label{as:fitting}

\subsection{Without a glitch}\label{ass:noglitch_fitting}

As we pointed out in \citet{lit:vanLier2025}, there are several difficulties when it comes to extracting asymptotic mixed mode parameters from a fit to Eq.\,(\ref{eq:Pi_as_mixed}). While we use an observation-like approach to the problem to simulate how the parameters would be obtained from observations, there are many advantages to using model frequencies, which we still utilize to optimize our results. These advantages consist mainly in knowing the radial orders $n_\mathrm{pg}$ and $n_\mathrm{p}$ of each mode\footnote{In order to allow for potential miss-assignments by GYRE, we redefined $n_\mathrm{p}$ (see \citet{lit:vanLier2025} for details) and extended the range of $\Phi$ we search beyond $[0,1]$.} and not being limited to p-dominated modes. We tested that restricting our analysis to p-dominated modes did only increase the uncertainties on some parameters but did not significantly change the values obtained. Still, even with the full mode sample the distribution of $\chi^2$~values across parameter space is complex with multiple local minima and strong correlations of the parameters, which makes numerical optimization difficult.

\begin{figure}[ht]\centering
    \resizebox{0.87\linewidth}{!}{\includegraphics{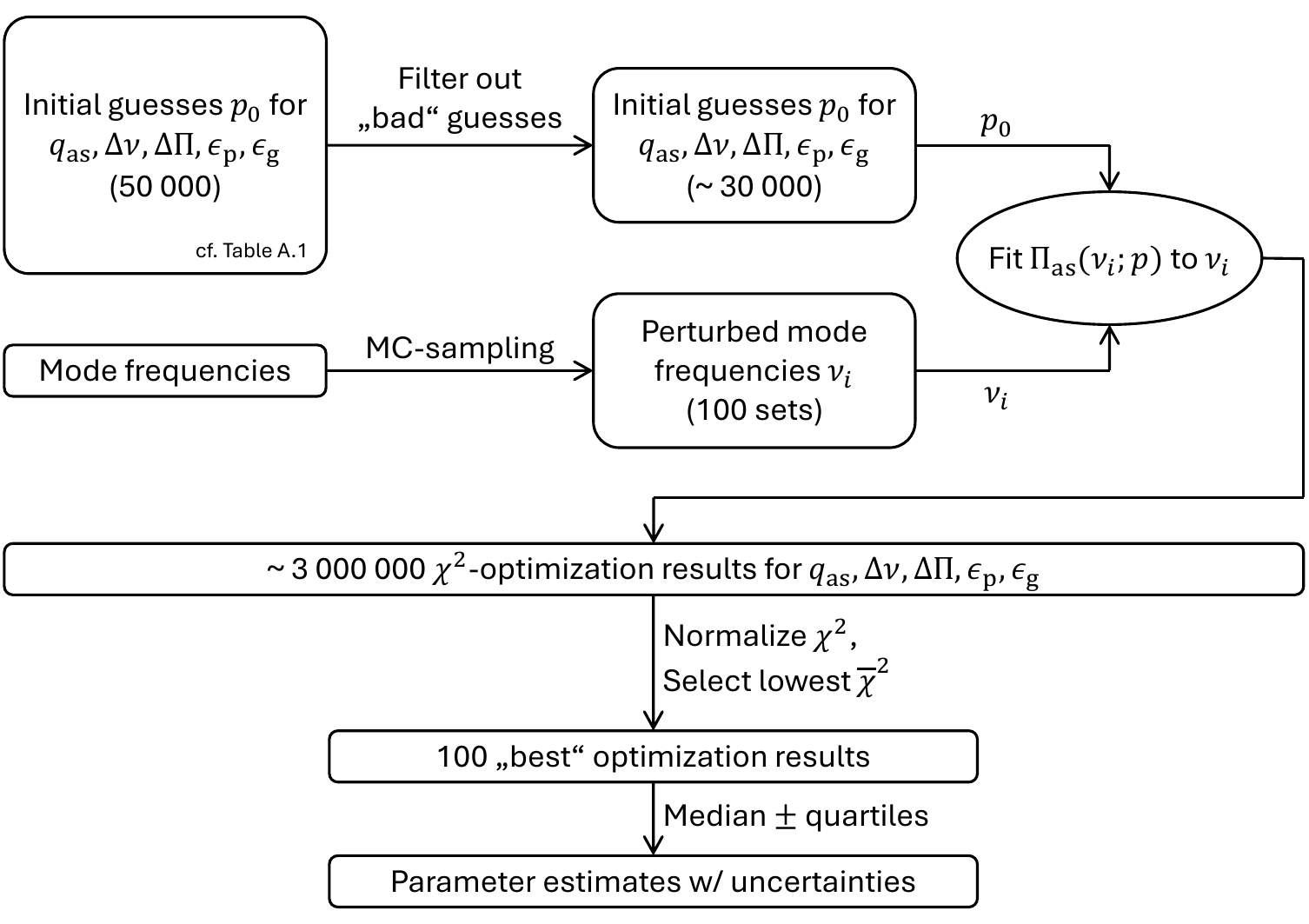}}
    \caption{Flowchart outlining our fitting procedure without taking glitches into account.}
    \label{fig:flowchart_noglitch_fitting}
\end{figure}

For each model, we therefore followed the algorithm outlined in Fig.\,\ref{fig:flowchart_noglitch_fitting}. To make sure we sample all local minima, we generated 50\,000 sets of initial guesses for the parameters, which we drew randomly from the distributions specified in Table~\ref{tab:parameter_shooting_noglitch_fitting}. To reduce computation time, we filtered out sets that produced a high initial $\chi^2$~sum, suggesting that they lie far away from the region of convergence. To simulate observational noise (and to prevent chance alignments from interfering with the statistics) we also applied a Monte-Carlo sampling to the mode frequencies observed for the model. This left us with roughly 3~Million combinations of perturbed mode frequencies and initial guesses for the parameters. For each of these combinations we then performed a numerical $\chi^2$~optimization and saved the optimized parameters and corresponding $\chi^2$~sum upon convergence. Once again to ensure that all local minima are sampled, we left the convergence criteria rather loose. A side effect of this choice was that we observed many limit convergences for the coupling parameter $q$, which has a hard physical limit at $q=0$. To reduce their impact on the statistics of our optimization results, we discarded all optimizations that converged to this limit. After this, we selected the results of the 100 optimizations that converged to the lowest $\chi^2$~values and calculated the median and quartiles of each parameter among these to get our final estimate for the parameter value and uncertainty.

\begin{table}[t]\footnotesize
    \caption{Properties of the initial-guess sampling without a glitch}
    \centering
    \begin{tabular}{llll}
        \hline\hline Parameter&Estimate&Width&Distribution\\\hline
        $q$ & 0.2 & 0.2 & Gaussian \\
        $\Delta\nu$ & $\Delta\nu_0$ ($l=0$ fit) & $0.03\cdot\Delta\nu_0$ & Gaussian \\
        $\Delta\Pi$ & $\Delta\Pi_N$ ($N$ integral) & $0.03\cdot\Delta\Pi_N$ & Gaussian \\
        $\varepsilon_\mathrm{p}$ & $\varepsilon_{\mathrm{p},0}+1/2$ & 0.7 & uniform \\
        & ($\varepsilon_{\mathrm{p},0}$: $l=0$ fit) & & \\
        $\varepsilon_\mathrm{g}=\Phi-\frac{1}{2}$ & $-0.7$ & 1.7 & uniform\\\hline
    \end{tabular}
    \title{}
    \tablefoot{For Gaussian distributions, the width refers to the standard deviation of the Gaussian; for uniform distributions, the limits were set to estimate$\pm$width. Since $q$ can physically only take positive values, all instances in which the Gaussian sampling returned a value $<0$ were resampled with a uniform distribution over $[0,0.8]$.}
    \label{tab:parameter_shooting_noglitch_fitting}
\end{table}

\subsection{With a glitch}\label{ass:glitch_fitting}

In order to explicitly treat a glitch as discussed in Sect.\,\ref{ss:glitch_cavity} or~\ref{ss:glitch_GaussCunha} in the fit, we included the different versions of the frequency-dependent phase $\Phi_\mathrm{cavity}$ into the asymptotic formula Eq.\,(\ref{eq:Pi_as_mixed}) to isolate the value for $\varepsilon_\mathrm{g}$:
\begin{align}
    \Phi=\varepsilon_\mathrm{g,as}+\frac{1}{2}-\Phi_\mathrm{cavity}\,.
\label{eq:Phi_w_glitch}
\end{align}
We then followed the fitting algorithm outlined in Fig.\,\ref{fig:flowchart_noglitch_fitting} again, using the new asymptotic periods including Eq.\,(\ref{eq:Phi_w_glitch}) and with a different sampling for the initial parameter guesses.

We found that the fit results for the asymptotic parameters $q$, $\Delta\nu$, $\Delta\Pi$, and $\varepsilon_\mathrm{p}$ are only weakly affected by including the glitch phase to $\Pi_\mathrm{as}$. Therefore, we used the values obtained from a fit as outlined in Appendix~\ref{ass:noglitch_fitting} to constrain the sampled distribution and increase the density of the initial guesses in parameter space. We restricted each of the four parameters to a range defined by the first and third quartile ($Q_1$ and $Q_3$) of the glitch-free best-fit distributions and a tolerance $\tau$ as
\begin{align}
	x \in \left[(1-\tau)Q_1(x),\,(1+\tau)Q_3(x)\right]\,.
\label{eq:tolerance_range}
\end{align}
The different tolerances for the various parameters (as given in Table~\ref{tab:glitchfit_parameter_limits}) ensured that the corresponding minimum of the $\chi^2$~sum including the glitch was still well sampled for each parameter in all models. When constructing the sets of initial parameters, we sampled $q$, $\Delta\nu$, $\Delta\Pi$ and $\varepsilon_\mathrm{p}$ uniformly across their respective allowed ranges. For $\varepsilon_\mathrm{g,as}$ we still used a uniform sampling over the range $[-0.7,1.7]$ as before.

\begin{table}[t]\footnotesize
    \caption{Constraints on parameters less affect by including the glitch}
    \centering
    \begin{tabular}{ll}
        \hline\hline Parameter&Tolerance $\tau$\\\hline
        $q$&10\,\%\\
        $\Delta\nu$&1\,\%\\
        $\Delta\Pi$&0.5\,\%\\
        $\varepsilon_\mathrm{p}$&2\,\%\\\hline
    \end{tabular}
    \title{}
    \label{tab:glitchfit_parameter_limits}
\end{table}

Since the glitch parameters $\{\mathfrak{C},\omega_\mathrm{g}^*\}$ for $\Phi_{\delta}$ and $\{C_\mathrm{G},\omega_\mathrm{g}^*,\Delta_\mathrm{g}\}$ for $\Phi_\mathrm{G}$ are (nearly) frequency-independent, we were able to treat them as free parameters in the fit. This is necessary for the Gaussian glitch to allow for corrections necessary due to the assumptions made in the derivation of $\Phi_\mathrm{G}$. For the Dirac-delta glitch we also observed a slight improvement of the fit results. To fit for the glitch parameters, we estimated a starting value from the stellar profile. Then we sampled the initial guesses from a Gaussian distribution around this starting value with a width of $5\,\%$. To avoid aliases due to the periodicity of the glitch phase in the glitch parameters, we restricted them to a range with width $25\,\%$ around the starting value during the fit.

\onecolumn
\section{Additional figures}

\subsection{Glitch-free phase}
\begin{figure*}[h!]
    \centering
    \includegraphics[width=17cm,trim={0 0 0 0.2cm},clip]{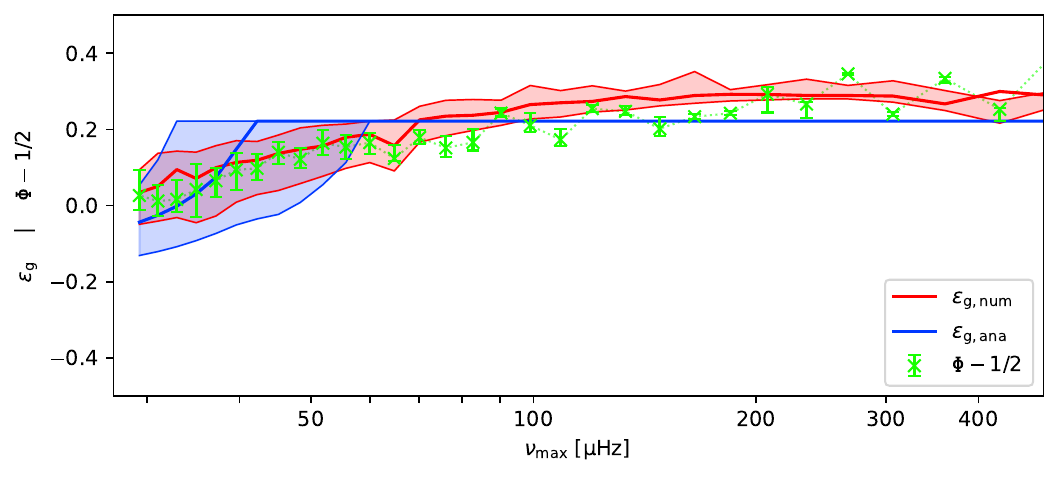}
	\caption{Same as Fig.\,\ref{fig:epsg_Pincon}, now for models along an evolutionary track where the mean--molecular weight discontinuity is artificially removed at each timestep.}
	\label{fig:epsg_nomudisc}
\end{figure*}

\subsection{Glitch parameters}
\begin{figure*}[h!]
    \centering
    \includegraphics[width=17cm]{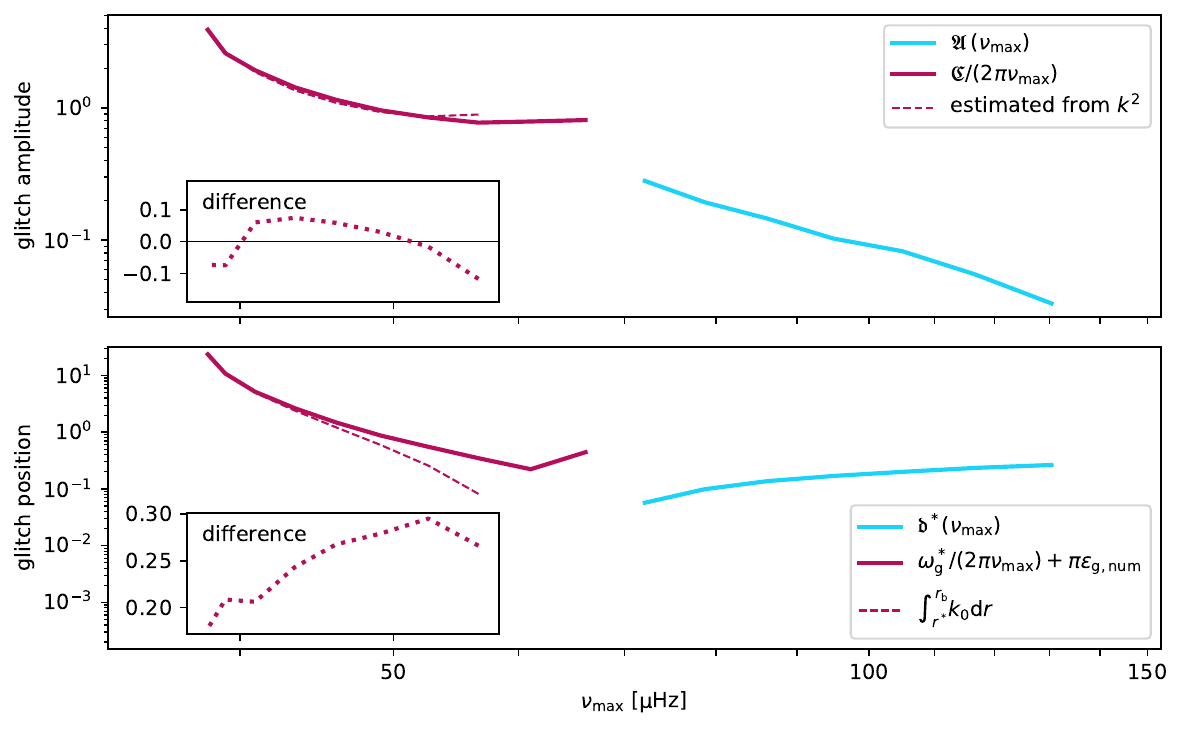}
	\caption{Glitch parameters of the models in Fig.\,\ref{fig:glitchfit_results} with a glitch in the evanescent zone (right) or buoyancy cavity (left) as a function of $\nu_\mathrm{max}$. The top panel shows dimensionless glitch amplitudes, the bottom panel shows dimensionless glitch positions. Stars evolve from right to left. Colors correspond to the colors of the fitting methods in Fig.\,\ref{fig:glitchfit_results} which use the respective glitch parameters. For comparison to the approximation in the limit $\omega^2\ll N^2,S^2$, the dashed lines show estimates for the parameters of a glitch in the cavity using the full asymptotic wave vector. Insets show the difference between the approximated glitch parameter and the estimate calculated using $k$.}
	\label{fig:glitch_parameters}
\end{figure*}

\newpage

\subsection{Mass and metallicity dependence}
\begin{figure*}[h!]
    \centering
    \includegraphics[width=17cm]{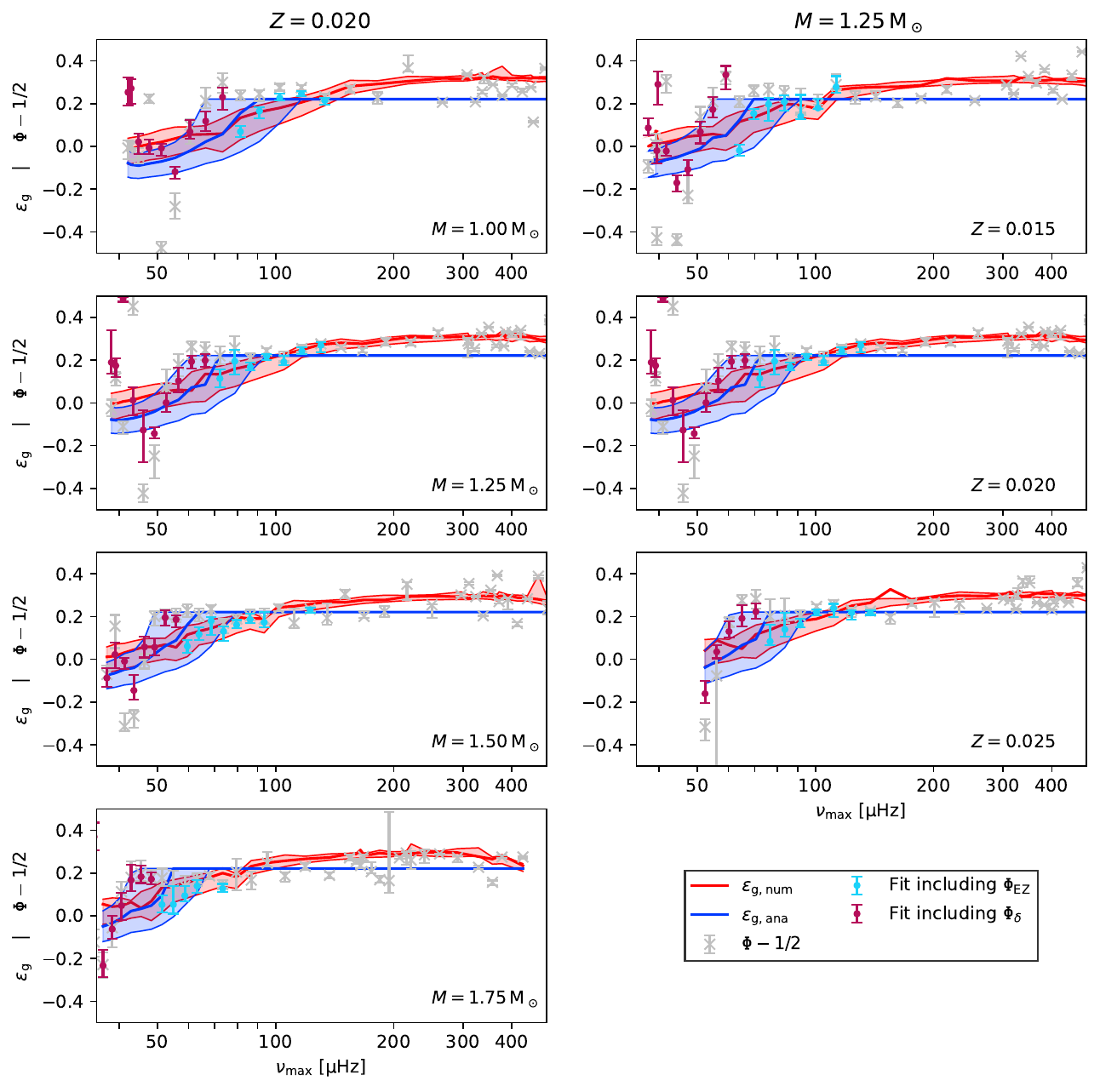}
    \caption{Same as Fig.\,\ref{fig:glitchfit_results}, now evaluated for stellar models marked in Fig.\,\ref{fig:HRD_selections} along the tracks with $Z=0.020,M\in\{1.00,1.25,1.50,1.75\}\,\mathrm{M}_\odot$ (left column, corresponding mass indicated in each panel) and $M=1.25\,\mathrm{M}_\odot,Z\in\{0.015,0.020,0.025\}$ (right column, corresponding metallicity indicated in each panel). For clarity, results of fits including $\Phi_\mathrm{G}$ are omitted as they show similar behavior as for $M=1.25\,\mathrm{M}_\odot,Z=0.020$.}
    \label{fig:epsg_MZ}
\end{figure*}

\end{appendix}

\end{document}